\documentclass[a4paper,10pt,twoside,bibnote]{cpc-hepnp}
\pdfoutput=1
\usepackage{CJK,upgreek,fancyhdr}
\usepackage{multicol}
\usepackage{graphicx}
\usepackage{booktabs}
\usepackage{slashed}
\usepackage{amssymb,bm,mathrsfs,bbm,amscd}
\usepackage[tbtags]{amsmath}
\usepackage{lastpage}
\usepackage{multirow}
\usepackage{textcomp}
\usepackage{epstopdf}
\usepackage{float}
\usepackage{footnote}
\usepackage{hyperref}
\usepackage{indentfirst}
\usepackage{subfigure}

\begin{document}
\begin{CJK*}{UTF8}{gbsn}
\setlength{\abovecaptionskip}{4pt plus1pt minus1pt}
\setlength{\belowcaptionskip}{4pt plus1pt minus1pt}
\setlength{\abovedisplayskip}{6pt plus1pt minus1pt}
\setlength{\belowdisplayskip}{6pt plus1pt minus1pt}
\addtolength{\thinmuskip}{-1mu}
\addtolength{\medmuskip}{-2mu}
\addtolength{\thickmuskip}{-2mu}
\setlength{\belowrulesep}{0pt}
\setlength{\aboverulesep}{0pt}
\setlength{\arraycolsep}{2pt}
\setlength{\parindent}{2em}

\title{Properties of secondary components in extensive air shower of cosmic rays in knee energy region} 
\author{Chen Yaling$^{1}$, Feng Zhang$^{1}$, Hu Liu$^{1,\dagger}$, Fengrong Zhu$^{1}$}

\maketitle

\footnotetext[2]{Email: huliu@swjtu.edu.cn}
\address{$^1$ School of Physical Science and Technology, Southwest Jiaotong University, Chengdu 610031, Sichuan, China}

\begin{abstract}
The “knee” of cosmic ray spectra reflects the maximum energy accelerated by galactic cosmic ray sources or the limit to the ability of galaxy to bind cosmic rays. The measuring of individual energy spectra is a crucial tool to ascertain the origin of the knee. However, the measuring of energy and the identifying of primary nuclei are the  foundation  of  measuring  the  energy  spectra  of  individual  components.  The  Extensive  Air  Shower  of cosmic rays in the knee energy region is simulated via CORSIKA software. The energy resolution for different
secondary  components  (include  electron,  gamma,  muon,  neutron  and  Cherenkov  light)  and  primary  nuclei
identification  capability  are  studied.  The  energy  reconstruction  by  using  electromagnetic  particles  (electron, gamma  and  Cherenkov  light)  in  the  energy  around “knee” is  better  than  by  using  other  secondary  particles. The  resolution  is  10\% –19\%  for  proton,  and  4\% –8\%  for  iron.  For  the  case  of  primary  nuclei  identification capability, the discriminability of density of muons is best both at low (~100 TeV) and high (~10 PeV) energy, the discriminability of the shape of lateral distribution of electron and gamma-rays are good at low energy and the  discriminability  of  density  of  neutrons  is  good  at  high  energy.  The  differences  between  the  lateral distributions  of  secondary  particles  simulated  by  EPOS-LHC  and  QGSJet-Ⅱ-04  hadronic  model  are  studied. For electron, gamma and Cherenkov light, the differences of the number of particles are within 5\%; for muon, when the perpendicular distance from the shower axis is greater than 100 m, the difference of the muon number is  within  5\%;  for  neutron,  the  difference  in  neutron  number  between  the  two  models  is  larger  than  10\%.  The results  in  this  work  can  provide  important  information  for  selecting  the  secondary  components  and  detector type during energy reconstruction and identifying the primary nuclei of cosmic rays in the knee region.
\end{abstract} 


\section{Introduction}
Cosmic ray is a high-energy particle from cosmic space, whose energy spectrum obeys the power law spectrum, and the maximum energy reaches about 1021eV .The main feature of its energy spectrum is that at about 1015eV. And the Spectral index of the power law spectrum changes from – 2.7 to – 3.1, which is called the "knee" region. The origin of cosmic ray knee region is an important subject in cosmic ray physics  \cite{1} . The “knee” of cosmic ray spectra reflects the maximum energy accelerated by galactic cosmic ray sources or the limit to the ability of galaxy to bind cosmic rays. Different models predict the different characteristics of the inflection energy (the energy where the Spectral index changes) of the single component cosmic ray energy spectrum in the knee region. For example, some models predict that the inflection energy is proportional to the charge Z of the original particle  \cite{2}, and some models predict that the inflection energy is proportional to the mass number A of the original particle \cite{3}.  The measurement of single component energy spectrum of cosmic ray is of great significance to the study of these transformations. 
\par At present, cosmic ray can be measured directly or indirectly. Direct measurement is mainly to measure cosmic ray through High-altitude balloon and space experiments, such as CREAM  \cite{4}, AMS  \cite{5}, DAMPE  \cite{6,7,8}, etc. Its advantage is that it can directly measure the charge of primary particles, and it has good discrimination ability for cosmic ray with different charges. At the same time, it can use the beam of accelerator experiment to calibrate the detector, and the absolute energy scale is relatively easy to determine. However, due to load limitations, the effective detection area is small, and the upper limit for measuring the energy spectrum can only reach around 100 TeV  \cite{9}. Therefore, the measurement of cosmic ray in the knee area mainly relies on indirect measurements from ground-based experiments, such as KASCADE \cite{10}, ARGO-YBJ \cite{11}, LHAASO \cite{12}, ICECUBE \cite{13}, TALE \cite{14}, TUNKA \cite{1} and AS-g \cite{15} . The ground experiment measures the primary cosmic ray by measuring the secondary components produced by the cosmic ray in the extensive air shower (EAS). Compared with the direct measurement method, it has the advantage of large effective detection area and can measure the energy spectrum of cosmic ray in the knee region. However, because the original cosmic ray particles are not directly measured, the ability to identify the composition of cosmic ray is not high, and the method of energy reconstruction often depends on the composition of the original particles and the absolute energy scale is not easy to determine, so for ground experiments, the energy measurement and the ability to identify the composition of the original Cosmic ray are the constraints for accurate measurement of single component energy spectrum.
\par
At present, most experiments only measure one or more of the secondary particles. For example, the detection energy band of KASCADE/KASCADE-Grand experiment is about 100 TeV-100 PeV, which can detect the electron, muon and hadron components in the secondary particles \cite{10}. The proton, helium nucleus, carbon, silicon and iron elements in the "knee" cosmic ray are identified and measured by the electromagnetic particle number and muon number \cite{16,17}; ARGO-YBJ and LHAASO-WFCTA prototypes detected charged particles and Cherenkov photons in secondary particles, and measured the full particle energy spectrum and light component energy spectrum of cosmic ray in the energy range of 1 TeV-10 PeV \cite{18}. The measurement energy band of ICETOP/ICECUBE is about 250 TeV-1 EeV \cite{19,20}, Aartsen et al\cite{20} used the deep learning technology to reconstruct the energy and composition of cosmic ray using Cherenkov photons generated by secondary particles in ice, thus realizing the energy spectrum measurement of components. These experiments measure different types of secondary components and the measurement results of energy spectra do not match \cite{16,17,18,19,20} . In this paper, we will study the energy reconstruction accuracy and particle identification ability of these secondary components, as well as their dependence on the strong interaction model. To provide reference for understanding the differences in measurement results between different experiments and how to obtain better energy reconstruction accuracy and particle identification ability.
\par
The measurement is conducted at the altitude where the longitudinal development of EAS reaches the maximum. The fluctuation of secondary particles is smaller, which can obtain better detection performance. Many experiments also measure cosmic ray at this altitude. In this paper, the secondary particles and Cherenkov photons of vertically incident cosmic ray at 4400 m above sea level will be studied in the knee region. Section \ref{sec:2} introduces the parameter settings for simulation, including the selection of detection planes, the parameter settings for secondary particles and Cherenkov light; Section \ref{sec:3} studies the horizontal distribution characteristics of secondary components in EAS and the differences between different Strong interaction models; Section \ref{sec:4} studies the energy reconstruction accuracy of secondary components to the original Cosmic ray; Section \ref{sec:5} studies the component identification ability of secondary components to the original Cosmic ray; Section \ref{sec:6} is a summary.
\section{EAS simulation}
\label{sec:2}
In this paper, CORSIKA Version-7.7410 software package \cite{21} is used to simulate the EAS of Cosmic ray in the atmosphere. EPOS-LHC and QGSJet - Ⅱ -04 are used as the high-energy strong interaction models, and EPOS-LHC Strong interaction model is used specifically. However, these two high-energy Strong interaction models are compared (figure \ref{fig:figure7} and figure \ref{fig:figure8}). The low-energy Strong interaction models use FLUKA, and the electromagnetic interaction models use EGS4. The five primary components are proton, helium, CNO, MgAlSi, and iron, respectively. The mass number of CNO and MgAlSi are 14 and 27, respectively. The initial particle energy $log_{10}(E/GeV)$ is fixed at 5.1, 5.3, 5.5, 5.7, 5.9, 6.1, 6.5, and 6.9. The zenith angle is fixed at 0 °, and the azimuth angle is evenly projected within 0 ° -360 °. In order to study the effects of non-vertical incidence, this article simulated a case where the zenith angle was fixed at 45 °, and compared the results of vertical incidence and zenith angle at 45 ° (figure \ref{fig:figure17}), with all other results being vertical incidence. The observation plane is selected at an altitude of 4400 meters, and the horizontal component and vertical component of the Earth's magnetic field at the observation plane are 34.618 $\mu T$ and   36.13 $\mu T$, respectively.
\par
The truncation kinetic energy of secondary particles is set to: hadron 0.1 GeV, muon 0.1 GeV, electron 1 MeV, and gamma ray 1 MeV. The selected truncation kinetic energy is lower than the default value in the CORSIKA manual to store more secondary particles, and the contribution of secondary particles below the selected truncation kinetic energy to the overall transverse distribution is very small. The wavelength of Cherenkov light is set at 200-1000 nm. The collection area of Cherenkov photons is a circular area whose vertical distances from the shower axis are respectively r=20, 50,100, 150, 200, 300 and 400 m. The radius of the circle is 3 m. In real experiments, the atmosphere has absorption and scattering effects on Cherenkov light, including Rayleigh scattering, aerosol scattering and ozone absorption. But they depend on specific models, and this article mainly studies the detection performance under ideal conditions, so these processes will not be considered in the simulation for the time being.
\section{Lateral distribution of secondaries}
\label{sec:3}
\subsection{Type of secondaries}
\begin{figure}[htbp]
    \centering
    \includegraphics[width=0.5\textwidth]{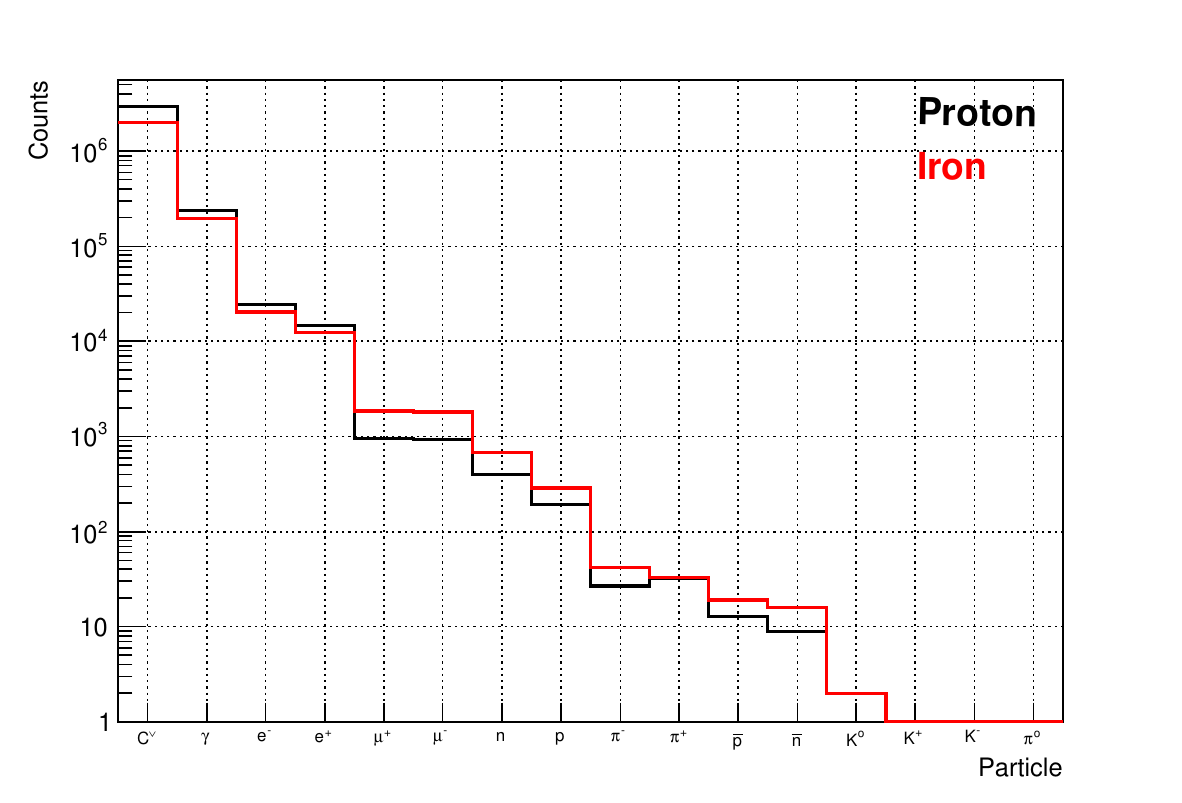}
    \caption{Type and counts of secondary components in the
EAS, the primary particles are proton (black) and iron
(red). Energy of the primary particle is $log_{10}(E/GeV)=5.1$.}
    \label{fig:fig1}
\end{figure}

The secondary components produced in EAS include Cherenkov photons, positron-electron , gamma rays, muons, neutrons, and other particles. Figure \ref{fig:fig1} shows the types and particle number of secondary components produced by Cosmic ray with protons (black) and iron nuclei (red) as their primary particles in the EAS process when energy is $log_{10}(E/GeV)=5.1$. Other primary cosmic rays produce similar secondary components in EAS, which will not be described here. Furthermore, in the setting observation plane, the secondary particles in descending order with a large number are Cherenkov photons, gamma rays, positron-electron, muon and neutrons. At present, most experiments are also conducted to measure these secondary particles, and this paper will only study these secondary components.
\subsection{Lateral distribution}
\begin{figure}[htbp]
    \centering
    \subfigure
     { \centering
      \label{fig:subfig2_a}
      \includegraphics[scale=0.3]{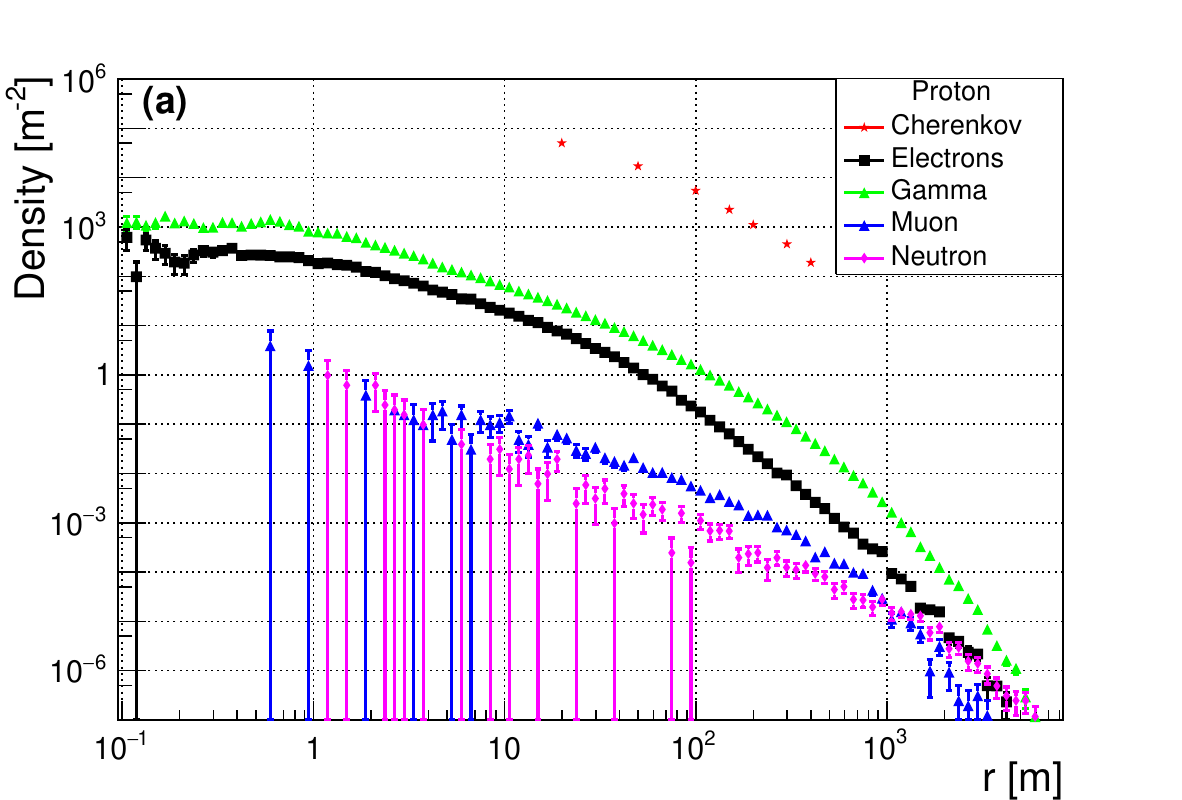}}
    \subfigure
    {  \centering
      \label{fig:subfig2_b}
      \includegraphics[scale=0.3]{image/fig2a.pdf}}
     \caption{Lateral distribution of secondary components produced by different primary particles during EAS: (a) Primary particle is proton with energy $log_{10}(E/GeV)=5.1$; (b) primary particle is iron with energy $log_{10}(E/GeV)=6.9$.}
 \label{fig:subfig2}
\end{figure}

In the EAS process, the vertical distance from the shower axis is recorded as r, and the relationship between the secondary particle number density at different locations and the location of r is the horizontal distribution of secondary particles. Figure \ref{fig:subfig2_a} and figure \ref{fig:subfig2_b} respectively show the transverse distribution of secondary components generated in EAS by Cosmic ray with energy $log_{10} (E/GeV)=5.1$, proton and energy $log_{10} (E/GeV)=6.9$, and iron core. It can be seen that at the same place, the number density of Cerenkov photons are thousands of times the number density of gamma ray photons, and the number density of gamma ray photons is 100-1000 times the density of neutron particle number with the smallest density. With r=100m and energy $log_{10}(E/GeV)=6.9$, the primary particle takes iron core as an example. Cherenkov photons number density is about $4×10 m^{– 2}$, the gamma ray density is about $100 m^{-2}$, the positron-electron number density is about $10 m^{-2}$, the muon number density is about $0.3 m^{-2}$ and the neutron number density is about $0.05 m^{-2}$. 
\par
\begin{figure}[htbp]
    \centering
    
    \subfigure
      {\centering
        \label{fig:subfig3_a}
      \includegraphics[scale=0.3]{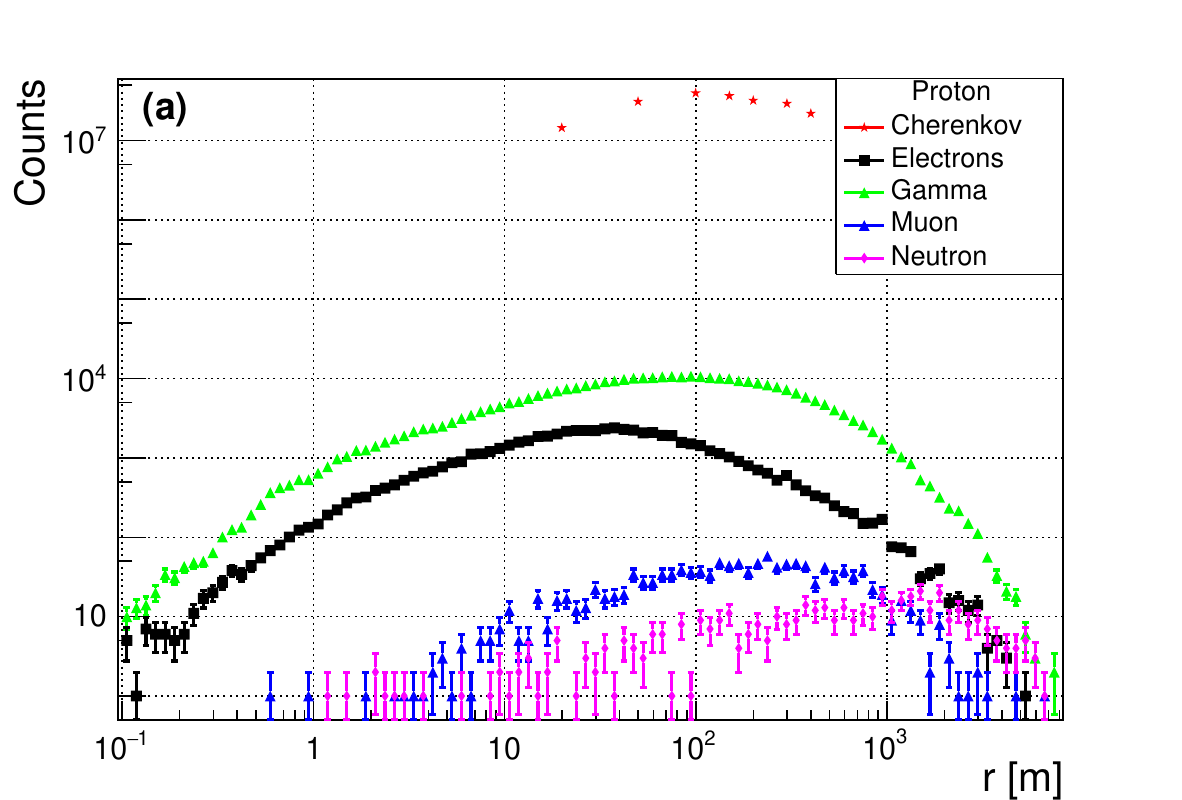}}
    \subfigure
      {\centering
        \label{fig:subfig3_b}
      \includegraphics[scale=0.3]{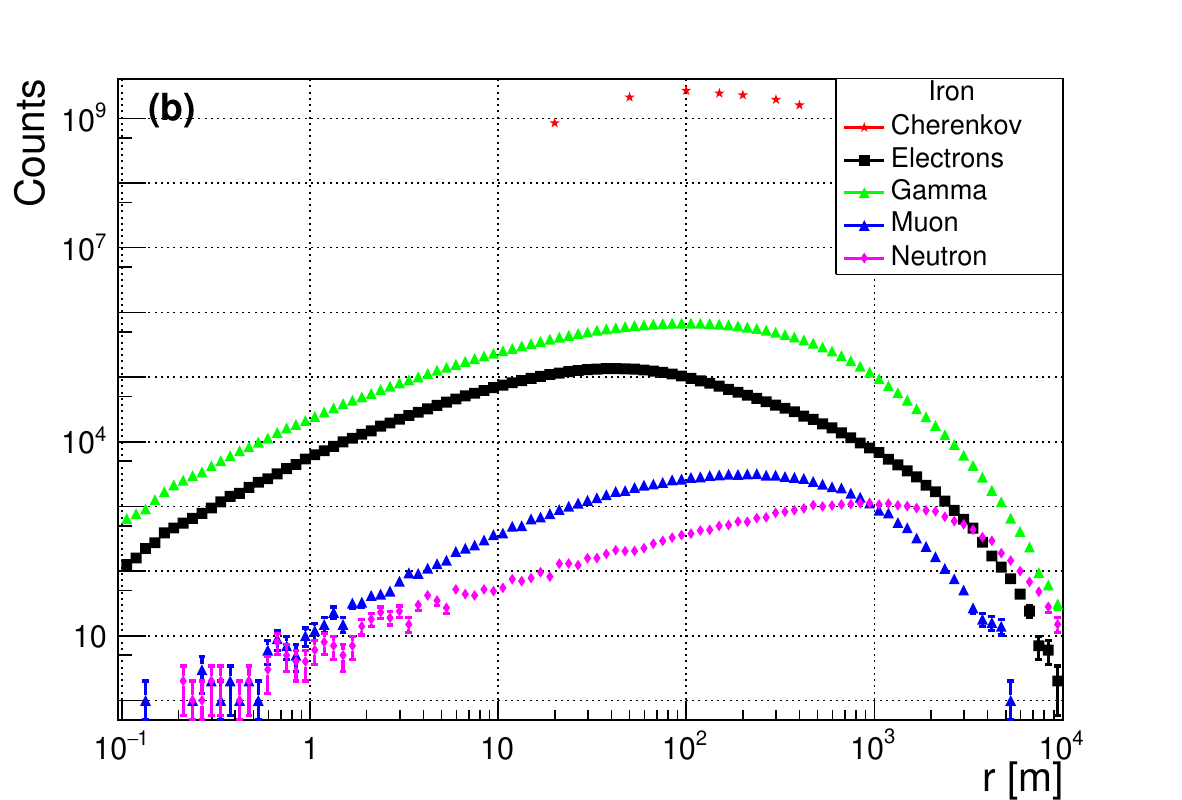}}
    \caption{Distribution of the number of secondary components produced by different primary particles during EAS in the detection
plane: (a) Primary particle is proton with energy $log_{10}(E/GeV)=5.1$;  (b) primary particle is iron with energy $log_{10}(E/GeV)=6.9$.}
\label{fig:figure3}
\end{figure}

In order to facilitate the viewing of the distribution range of secondary particles in the detection plane, a ring is taken with the core as the center and the vertical distance from the core as the radius. The radius of the ring is taken and the number of secondary components in the ring is counted, as shown in figure \ref{fig:figure3}. Figure \ref{fig:subfig3_a} and figure \ref{fig:subfig3_b} respectively show the primary particles with energy $log_{10}(E/GeV)= 5.1$, the primary particles with energy $log_{10}(E/GeV)=6.9$ composed of protons and the primary particles with iron core. The abscissa is a uniform ring for $log_{10}(r)$. It can be seen that for primary Cosmic ray with different energies and components, the number of secondary particle  first increases with the vaule of r. When it reaches the maximum, the number of secondary particle start to decrease with the vaule of r. Positive and negative electrons are mainly distributed within 10-100 meters; Gamma rays and muons are mainly distributed within the range of tens to hundreds of meters from the core site; Neutrons are mainly distributed around 1 km away from the core; and Cherenkov photons are mainly distributed near the 100 meters away from the core.
\par
For EAS generated by electromagnetic particles, the Kamata-Greisen (NKG)  function is usually used to describe the lateral distribution of its secondary particles, expressed as\[\rho_{1}(r)=N_{\mathrm{size}} C(s)\left(\frac{r}{R_{\mathrm{M}}}\right)^{s-2}\left(1+\frac{r}{R_{\mathrm{M}}}\right)^{s-4.5}\]
\begin{equation}
C(s)=\frac{1}{2 \pi R_{\mathrm{M}}^{2}} \times \frac{\Gamma(4.5-s)}{\Gamma(s) \Gamma(4.5-2 s)}
\label{eq1}
\end{equation} In the formula, C(s) is a function of s, $\Gamma$ represents the Gamma function, the vertical distance from the EAS shower axis, the particle population density at the location, the total number of secondary particles, the Morrill radius at the location of the observation plane, and the age of EAS development\cite{22}.
\par
For Cosmic ray whose primary particles are protons, helium, oxygen, silicon and iron, different ground experiments have modified the NKG function to describe the transverse distribution of its secondary particles. For example, in the KASCADE experiment, the expression of describing the lateral distribution of secondary particles produced by hadrons in EAS is \cite{23} \[\rho_{2}(r)=N_{\text {size }} C(\lambda)\left(\frac{r}{r_{0}}\right)^{\lambda-\alpha}\left(1+\frac{r}{r_{0}}\right)^{\lambda-\beta}\]
\begin{equation}
    C(s)=\frac{1}{2 \pi r_{0}^{2}} \times \frac{\Gamma(\beta-\lambda)}{\Gamma(\lambda-\alpha+2) \Gamma(\alpha+\beta-2 \lambda-2)}
    \label{eq2}
\end{equation}In the formula, the parameter represents the age of EAS development, which is a free parameter. But $r_0$, α and β are defined as a constant. For the KASCADE experiment, α = 1.5, β = 3.6 and $r_0$ = 40 meters.
\par
This article first attempts to use equation (\ref{eq2}) to fit the lateral distribution of different secondary components in figure \ref{fig:subfig2}. It was found that there can be multiple sets of fitting parameters for the same horizontal distribution, that is, there is coupling between the parameters (for example, only two parameters are independent in λ，α and β). In order to reduce fitting parameters, this article adopts a more general equation (\ref{eq3}) to fit the lateral distribution of secondary components:\[\rho(r)=N_{\mathrm{size}} C(s)\left(\frac{r}{r_{0}}\right)^{s-2}\left(1+\frac{r}{r_{0}}\right)^{(s+\Delta)}\]\begin{equation}
    C(s)=\frac{1}{2 \pi r_{0}^{2}} \times \frac{\Gamma(-s-\Delta)}{\Gamma(s) \Gamma(-\Delta-2 s)}
    \label{eq3}
\end{equation}
In the equation, $\Delta$ is the parameter. When  $\Delta$= -4.5 and $r_0$ = $R_M$ in equation (\ref{eq3}), it is consistent with equation (\ref{eq1}); When s=λ+0.5,  $\Delta$ = -4.1 and $r_0$ = 40 meters , it was consistent with equation (\ref{eq2}) . The formula (\ref{eq3}) is a double power law function. The specific meaning of the parameter is to represent the power law index (or slope) of the phase where the Particle number increases with the increase of the value of r in figure \ref{fig:figure3}, which is equivalent to the age parameter in Formula (\ref{eq1}). 2s+ $\Delta$ represents the power law index (or slope) of the phase where the Particle number decreases with the increase of the value of r in figure \ref{fig:figure3}, and $r_0$ represents the coordinates of r where the two different power law indexes change.
\par
\begin{figure}[htbp]
    \centering
    \subfigure
      {\centering
      \label{fig:subfig4_a}
      \includegraphics[scale=0.3]{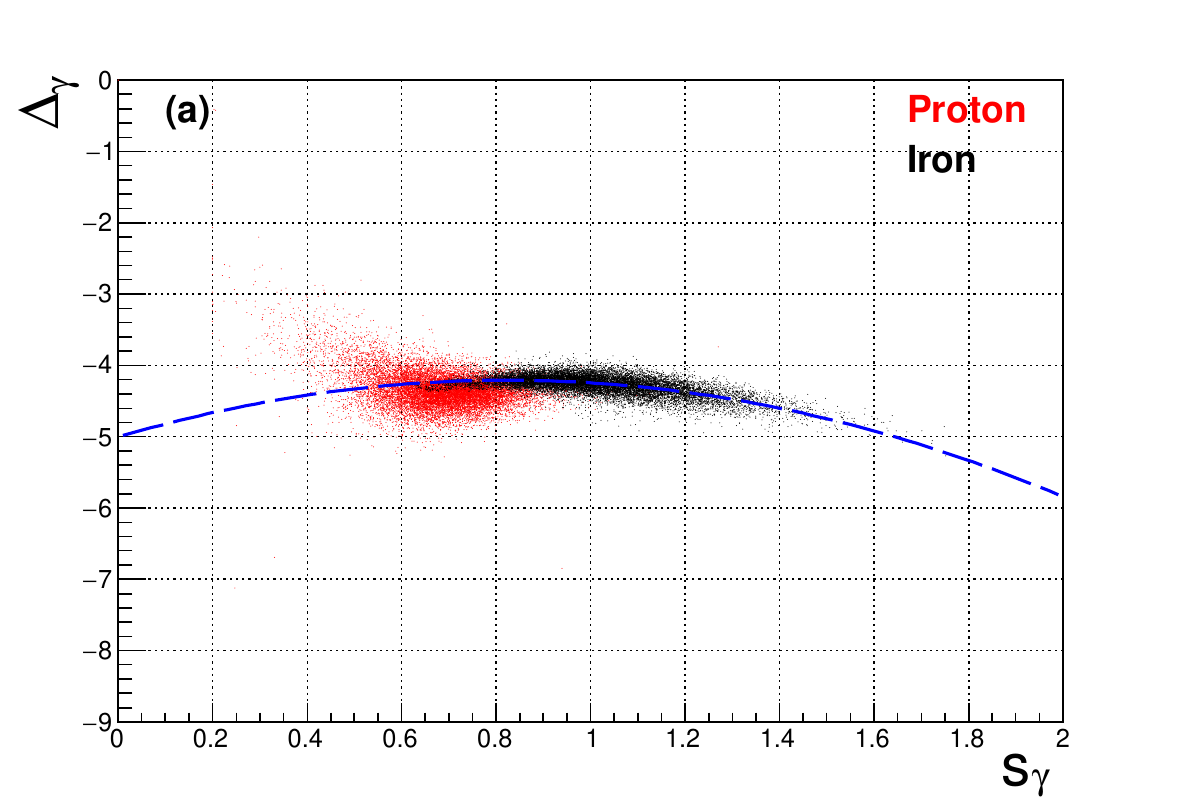}}
    \subfigure
      {\centering
      \label{fig:subfig4_b}
      \includegraphics[scale=0.3]{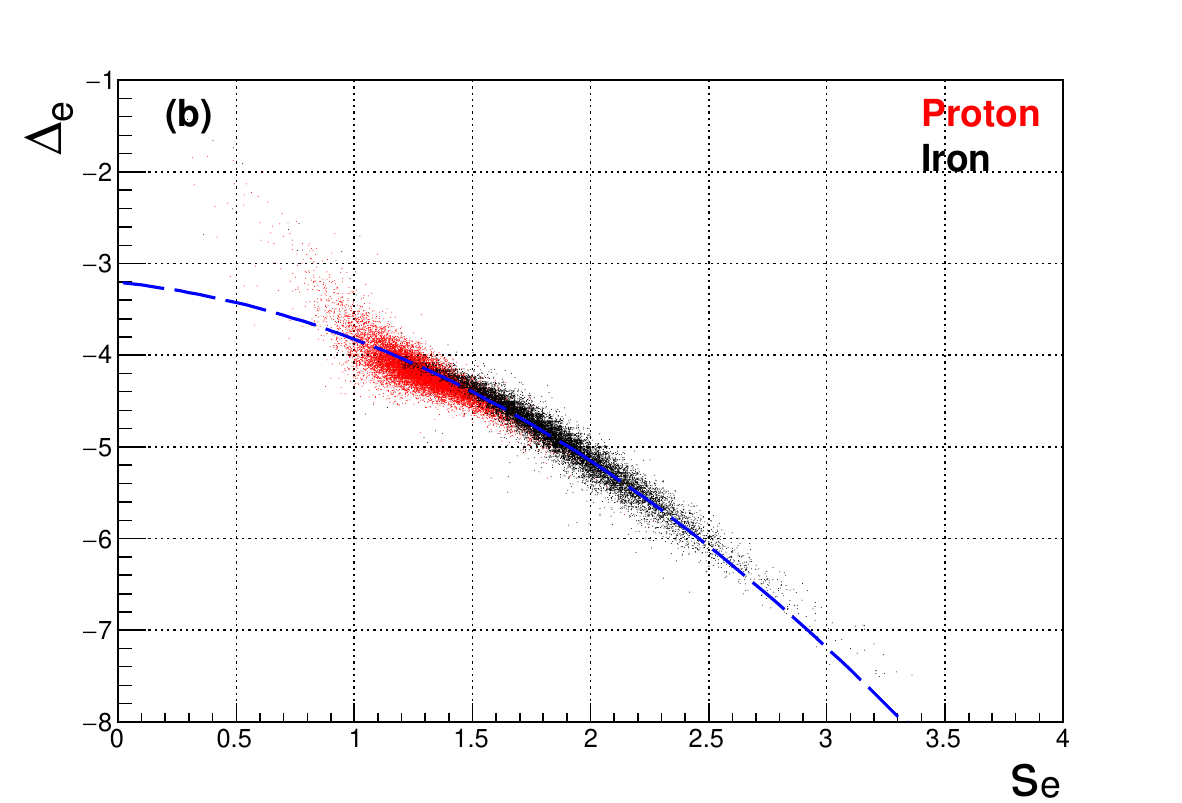}}
     \label{fig:figure4}
    \caption{Dependence between parameters s and $\Delta$
 in gamma (a) and electron (b) lateral distribution fitting (Shower is induced by
proton and iron respectively, and the blue dotted line is the fitting curve).}
\end{figure}

There are four free parameters in the formula (\ref{eq3}) that are $N_{size}$, s, $\Delta$,
$r_0$.  And the value of these fitting parameters are not unique. The same horizontal distribution can be fitted by combining multiple groups 
of parameters. Multiple parameter combinations can be used to fit the same lateral distribution. In order to further reduce the number of 
free parameters, the correlation between these parameters was
studied. For the energy segment simulated in this paper
and the original cosmic
ray component, when the secondary particle is
gamma ray and $r_0^γ=460m$, the horizontal
distribution of these
cases can be fitted. And the
correlation between $s_\gamma$ and $\Delta_\gamma$ is shown in figure \ref{fig:subfig4_a} and the
fitting expression is Formula (\ref{eq4}). So $N_{\text {size }}^{\gamma}$ ,$s_e$ will only
be used as free parameters. Correspondingly, for the electrons in the secondary particles,
when $r_{0}^{\mathrm{e}}=50$m, fitting $s_e$ and $\Delta_e$ satisfy equation (\ref{eq5}), as 
shown in figure \ref{fig:subfig4_b} $N_{\text {size }}^{\mathrm{e}}$ and $s_e$  will only be used as 
free parameters. For the muon in the secondary particle, 
fixed $r_{0}^{\mu}=800$m, the
$s_\mu$ and $\Delta_\mu$ satisfy equation (\ref{eq6}) and $N_{\text {size }}^{\mu}$,$s_\mu$ will only
be used as free parameters.
\begin{equation}
    \Delta_{\gamma}=-1.18 \cdot s_{\gamma}^{2}+1.94 \cdot s_{\gamma}-5.00
    \label{eq4}
\end{equation}
\begin{equation}
    \Delta_{\mathrm{e}}=-0.35 \cdot s_{\mathrm{e}}^{2}-0.27 \cdot s_{\mathrm{e}}-3.20
     \label{eq5}
\end{equation}
\begin{equation}
    \Delta_{\mu}=-s_{\mu}-4.4 
     \label{eq6}
\end{equation}

\par
For neutrons in secondary particles, due to the small number of secondary particles, the limit on the transverse distribution function is weaker. And the range of variation for each parameter is larger. According to equation (\ref{eq3}) in reference  \cite{24}, certain parameters of equation (\ref{eq3}) in this paper were fixed and optimized based on this. It was found that equation (\ref{eq7}) can fit the transverse distribution of neutrons well,where the free parameter is $N_{\text {size }}^{\mathrm{n}}$and $r_{0}^{\mathrm{n}}$.
\[\rho_{\mathrm{n}}(r)=N_{\mathrm{size}}^{\mathrm{n}} C_{\mathrm{n}}\left(\frac{r}{r_{0}^{\mathrm{n}}}\right)^{-0.9}\left(1+\frac{r}{r_{0}^{\mathrm{n}}}\right)^{-4.0}\]
\begin{equation}
   C_{\mathrm{n}}=\frac{1}{2 \pi\left(r_{0}^{\mathrm{n}}\right)^{2}} \times \frac{\Gamma(4.0)}{\Gamma(1.1) \Gamma(2.9)} 
    \label{eq7}
\end{equation}
\begin{figure}[htbp]
    \centering
    \subfigure
     { \centering
      \label{fig:subfig5_a}
      \includegraphics[scale=0.3]{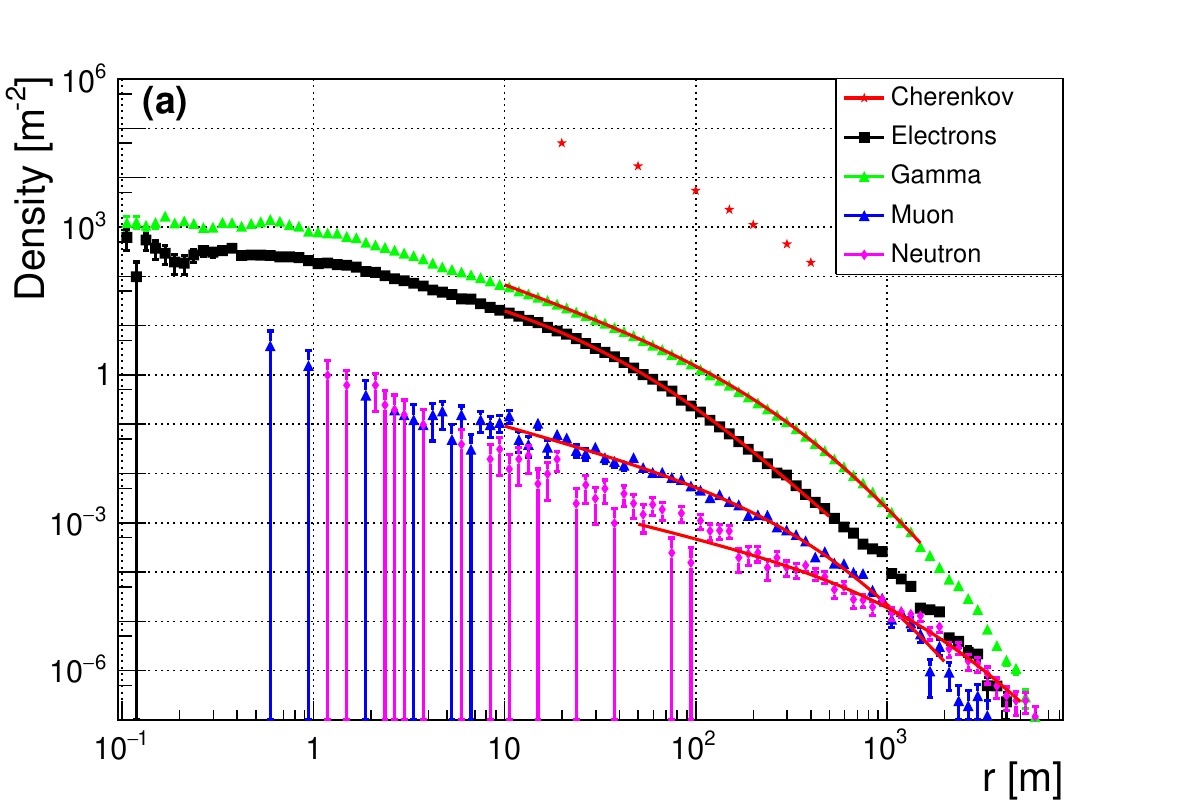}}
    \subfigure
      {\centering
      \label{fig:subfig5_b}
      \includegraphics[scale=0.3]{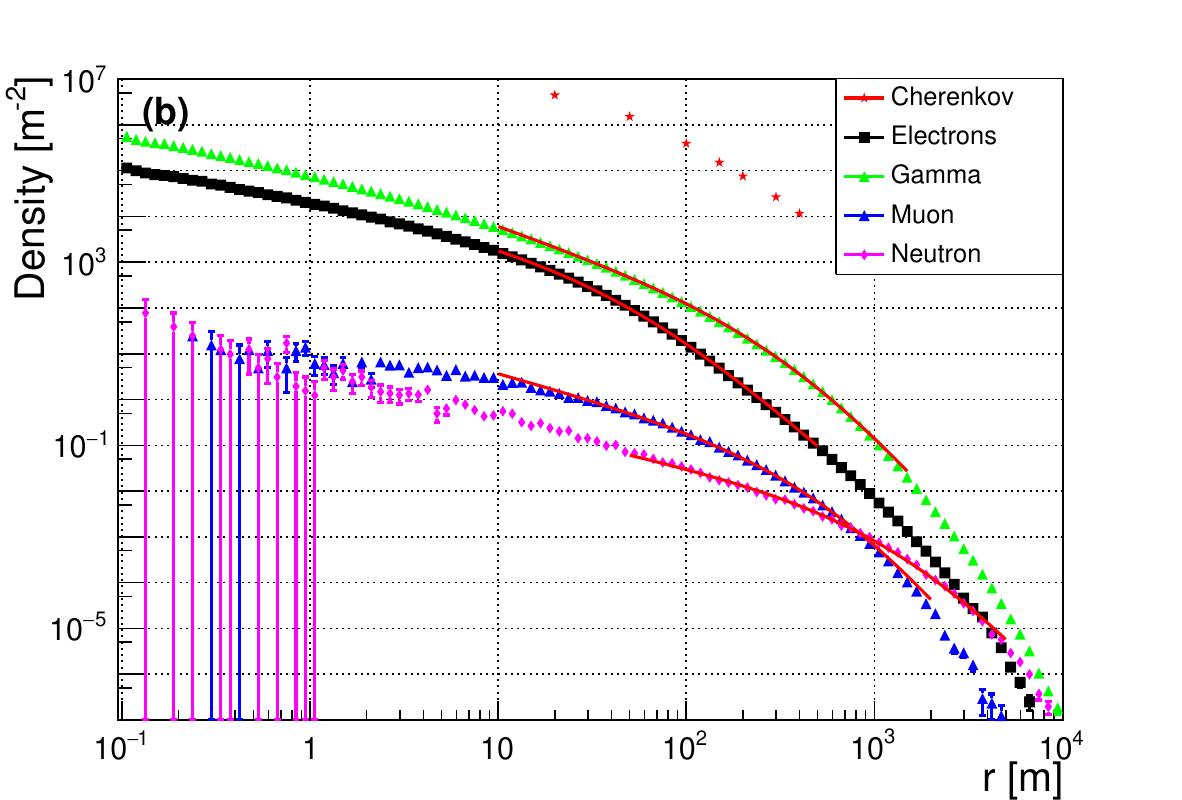}}
    \subfigure
      {\centering
      \label{fig:subfig5_c}
      \includegraphics[scale=0.3]{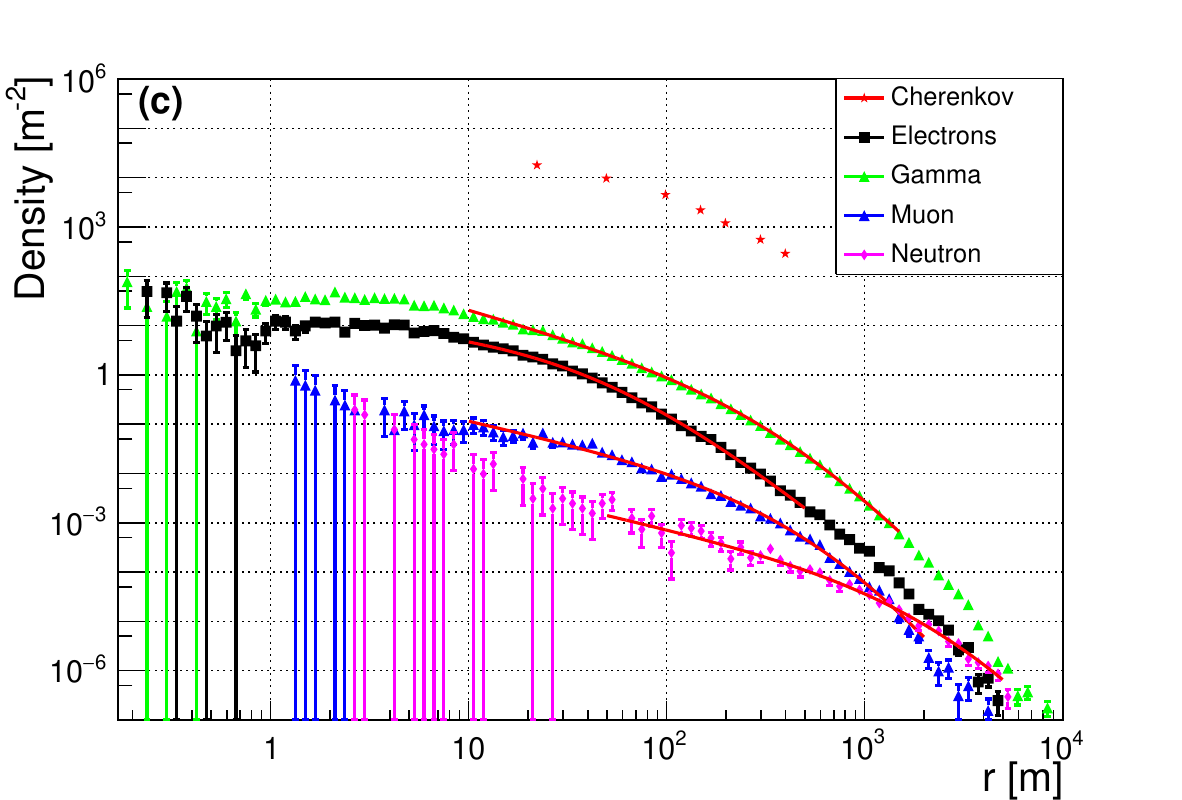}}
    \subfigure
     { \centering
      \label{fig:subfig5_d}
      \includegraphics[scale=0.3]{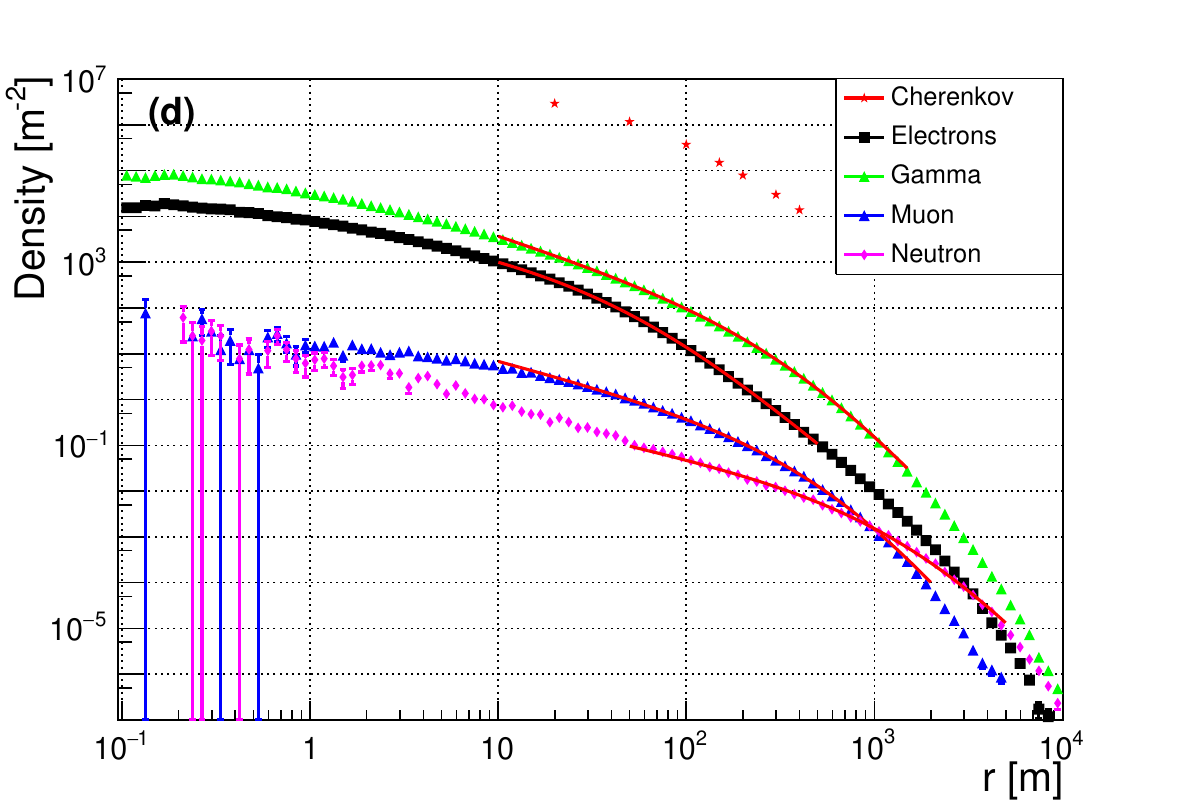}}
     \caption{Fitting of lateral distribution of secondary components: (a), (b) Primary particle is proton with $log_{10}(E/GeV)=5.1$(a)and $log_{10}(E/GeV)=6.9$ (b); (c),(d)  primary particle is iron with $log_{10}(E/GeV)=5.1$ (c) and $log_{10}(E/GeV)=6.9$ (d). The green, black,
blue, and pink points represent gamma, electron, muon, and neutron respectively, the red stars at the top represent Cherenkov
light. The solid lines with the same color are the fitted function.}
   \label{fig:figure5}
\end{figure}

\par
Equations (\ref{eq3}) and (\ref{eq7}) are used to fit the transverse distribution of secondary particles produced by cosmic rays of different components in EAS, as shown in figure \ref{fig:figure5}. Because only a few Cherenkov photons are preserved in the simulation, the transverse distribution of Cherenkov photons is not fitted.
\par
\begin{figure}[htbp]
    \centering
    \subfigure
     { \centering
     \label{fig:subfig6_a}
      \includegraphics[scale=0.3]{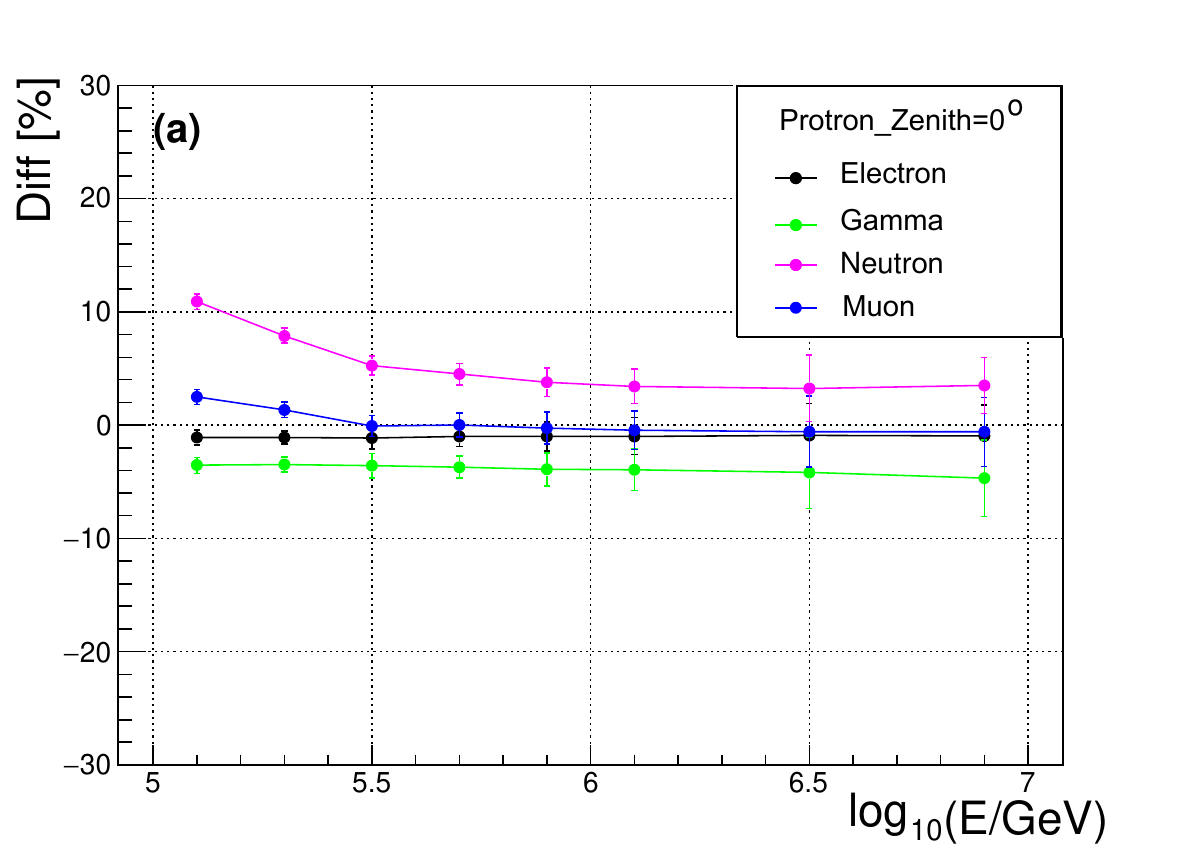}}
    \subfigure
     { \centering
      \label{fig:subfig6_a}
      \includegraphics[scale=0.3]{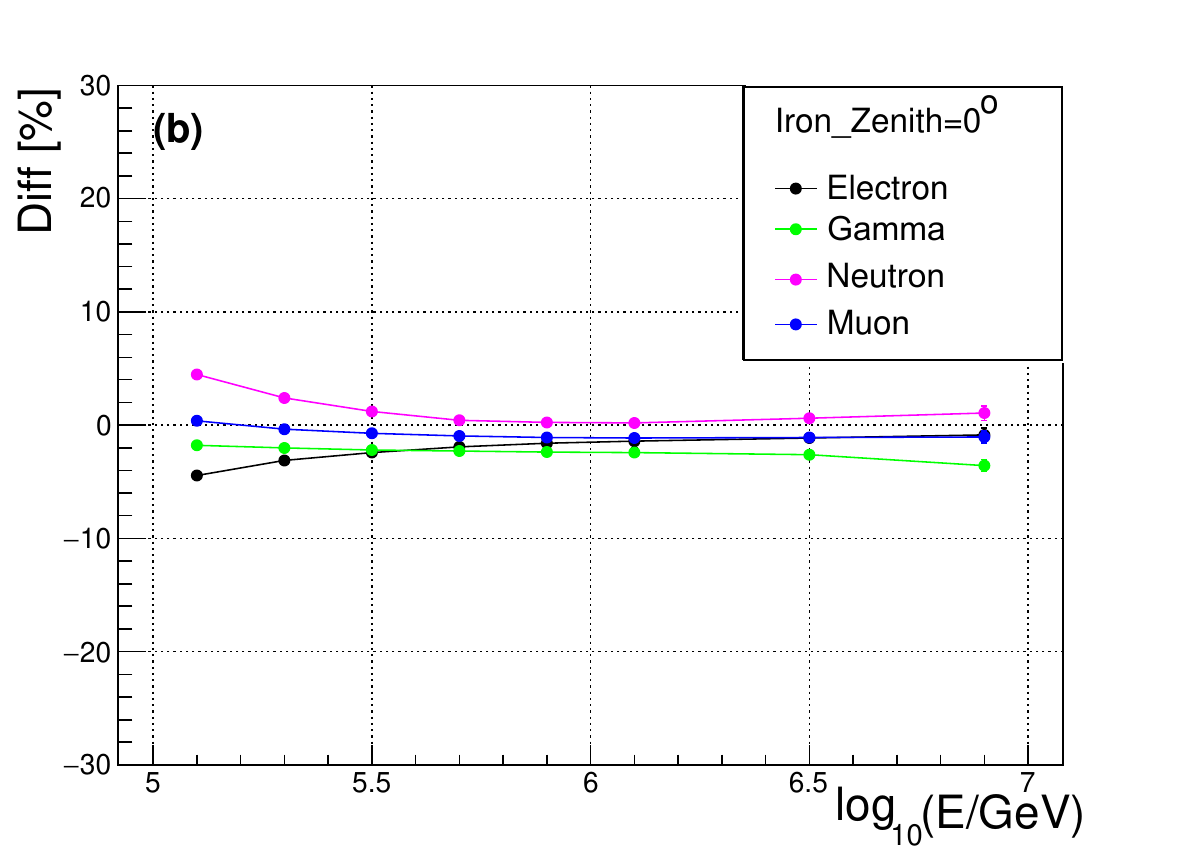}}
     \caption{Deviation between counted value N and fitted value
$N_{size}$
for different secondary particles. Shower is induced by proton (a)
and iron (b).}
 \label{fig:figure6}
\end{figure}

To test the fit quality of the transverse distribution of different secondary components, the deviation between the fit values “Nsize“and the statistical values “N”of the number of secondary particles produced by cosmic rays of different components and energies in EAS ($\text { Diff }=\frac{N-N_{\text {size }}}{N_{\text {size }}} \times 100 \%$) is shown in figure \ref{fig:figure6}. When the energy $log_{10}(E/GeV)>5.5$, the deviation of all particles is within 6\%, and then the fluctuation will be used to characterize the accuracy of the energy reconstruction.

\subsection{Differences between hadronic models}
\begin{figure}[htbp]
    \centering
    \subfigure
      \centering
      \includegraphics[scale=0.3]{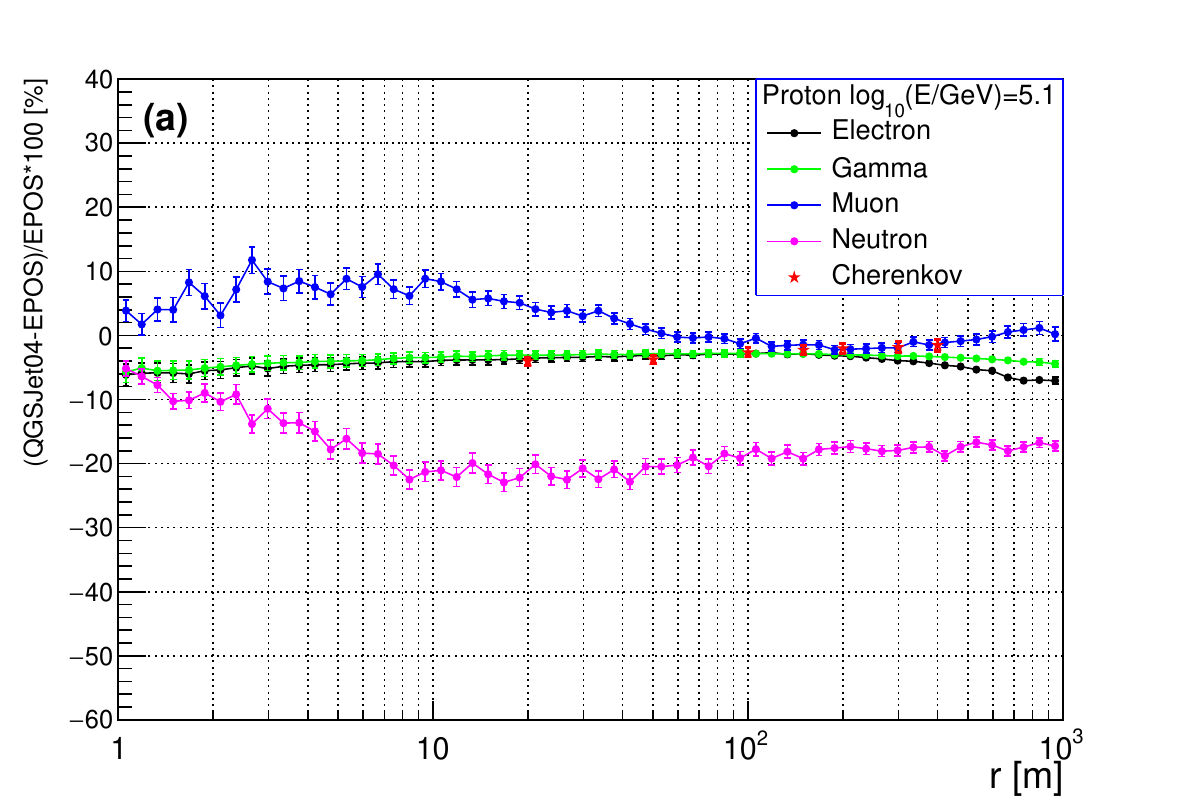}
    \subfigure
      \centering
      \includegraphics[scale=0.3]{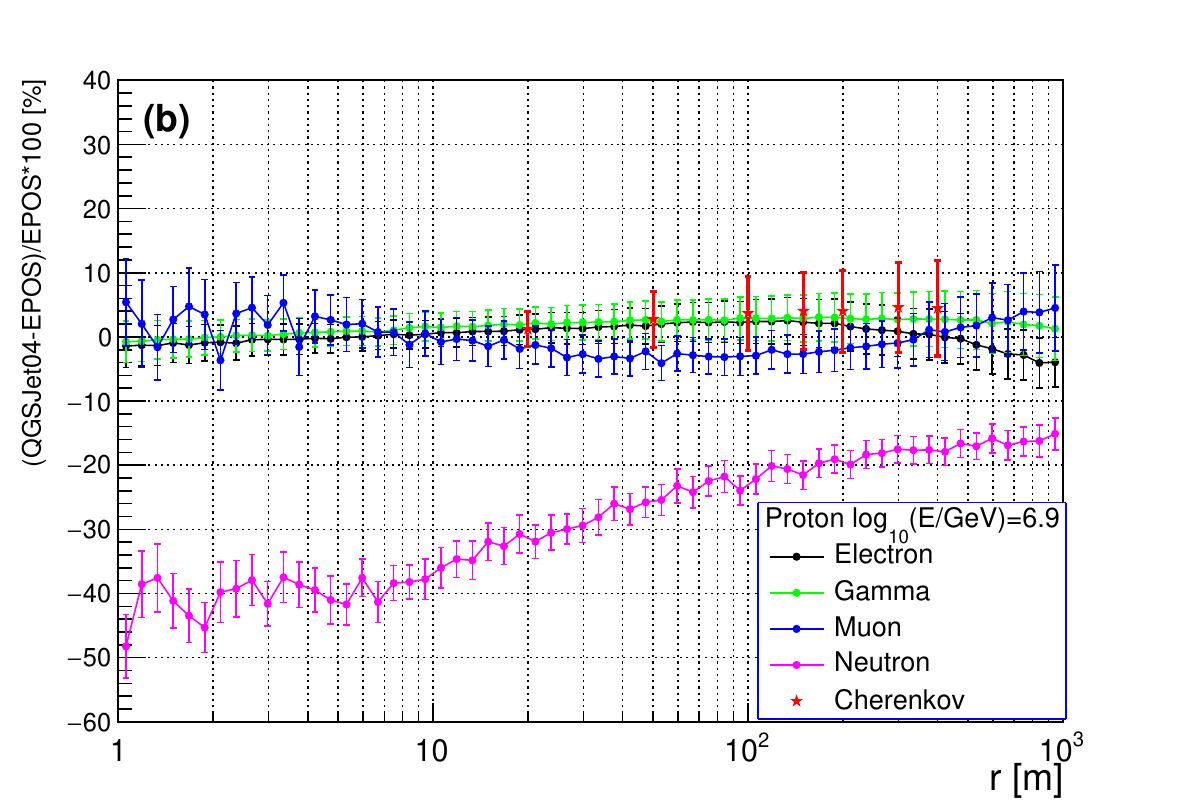}
    \caption{Difference in percentage of the lateral distribution of secondary particles between EPOS-LHC and QGSJet-Ⅱ-04 hadronic
interaction model, in which the primary particles are protons with different energies: (a) $log_{10}(E/GeV)=5.1$; (b) $log_{10}(E/GeV)=6.9$.
\label{fig:figure7}
}
\end{figure}

\begin{figure}[htbp]
    \centering
    \subfigure
      \centering
      \includegraphics[scale=0.3]{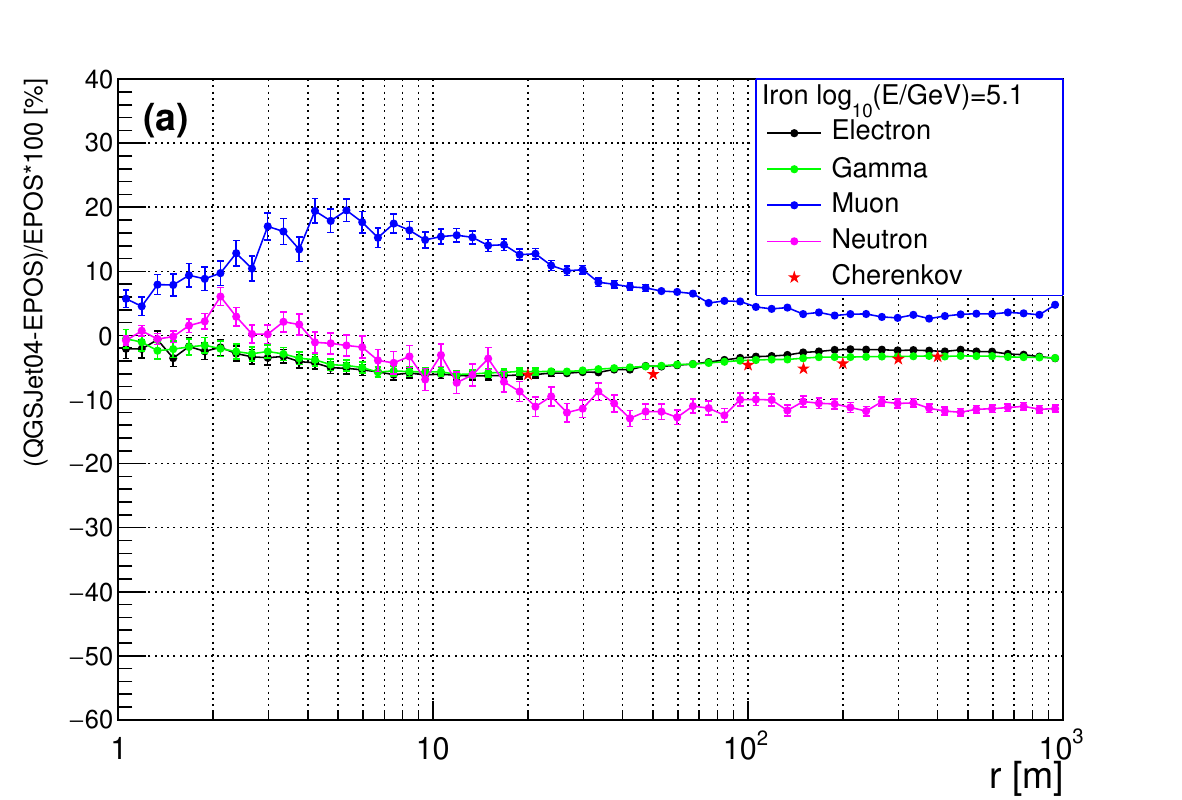}
    \subfigure
      \centering
      \includegraphics[scale=0.3]{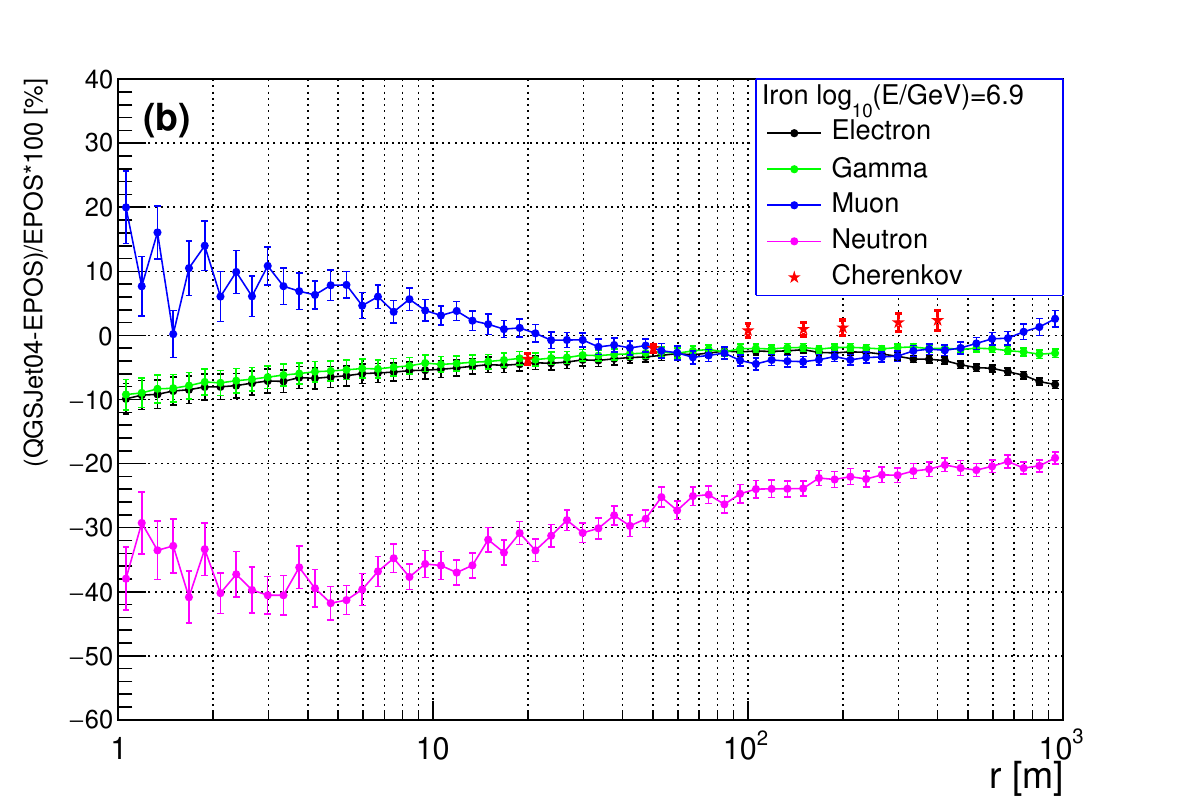}
    \caption{Difference in percentage of the lateral distribution of secondary particles between EPOS-LHC and QGSJet-Ⅱ-04 hadronic
interaction model, in which the primary particles are irons with different energies: (a) $log_{10}(E/GeV)=5.1$; (b) $log_{10}(E/GeV)=6.9$.}
\label{fig:figure8}
\end{figure}

The difference in current intensity of proton spectra in the knee region measured by different strong interaction models in KASCADE experiments is nearly double \cite{10}. In this paper, the two strong interaction models’s difference of transverse distribution between EPOS-LHC and QGSJet-Ⅱ-04 is studied, and the results are shown in figure \ref{fig:figure7} and figure \ref{fig:figure8}. It can be seen that the difference in the number of positrons, gamma rays and Cerenkov photons in the two models is close and minimal. When r$>$20m, the difference of the models of the above three particles is within 5\% and all range differences of r are within 10\%; when r$>$100m, the difference of Muon’s model is less than 5\%, but when r$<$100m, the maximum difference can be close to 20\% (corresponding to the original particle is iron, with an energy of about 10 PeV, around r= 5 m); Neutrons differ the most, with a difference of 10\%-20\% at r$>$100m and a maximum difference is about 40\% when r$<$100m (corresponding to r$<$100m). In general, the difference between muons and neutrons is significantly reduced when r$>$100m. For experiments measuring muse and neutrons, it is recommended that the detector size is greater than 100 m and should select particles greater than 100 m for reconstruction to reduce model dependence. Muons and neutrons are the products of the strong interaction process. The EPOS-LHC model takes into account the influence not considered in other strong interaction models. Under the multiple scattering of EPOS-LHC, the energy scale of a single scattering is taken into account when calculating the respective cross sections. This is not the case in the QGSJET-II-04 model based on Gribov-Regge theory \cite{25} .The differences between different strong interaction models have been studied in detail in literature  \cite{25,26,27} and will not be discussed in this paper.
\section{Energy resolution}
\label{sec:4}
By fitting the transverse distribution of the secondary particles produced by cosmic ray EAS, the fitting parameters of each secondary particle can be obtained. The number of particles or the number density of particles at a certain radius is often used for energy reconstruction, while the ratio of the number of different secondary particles and the shape parameters of the transverse distribution are often used to identify the composition of the original particle. In this paper, we will use the Cherenkov photon numbers of the four secondary particles fitted and statistically obtained at a different “r” in Section \ref{sec:3} to characterize the energy reconstruction accuracy and compare the differences between them. This paper only studies the accuracy of energy reconstruction under fixed components and does not study the component dependence of energy reconstruction and the construction of composition-independent energy reconstruction variables by combining real observation data with composition-sensitive variables, which is beyond the scope of this paper. The results obtained in this paper are superior to those obtained after taking into account the composition correction, so it can be considered as the upper limit of energy reconstruction using a single secondary particle. 
\par
Since the energy of the primary particle is proportional to the number or density of secondary particles,\begin{equation}
    E=C \times N_{\text {size }}
\end{equation}So the percentage spread of the number of secondary particles or the number density (defined as the spread of the number or number density distribution divided by the mean of the distribution) is equal to the resolution of the reconstructed energy\begin{equation}
    \frac{\Delta E}{E}=\frac{\Delta\left(C \times N_{\text {size }}\right)}{C \times N_{\text {size }}}=\frac{\Delta N_{\text {size }}}{N_{\text {size }}}
\end{equation} Because the simulation process is carried out at several discrete fixed energies, the influence of the width of the energy range on the broadening of the particle population distribution is not involved. Therefore, this paper will directly use the percentage of the distribution broadening of particle number or particle number density to characterize the energy reconstruction accuracy, without carrying out specific energy reconstruction, and the calculation of the broadening will use the value of $\sigma$ fitted by Gaussian function.
\par
\begin{figure}[htbp]
    \centering
    \includegraphics[width=0.5\textwidth]{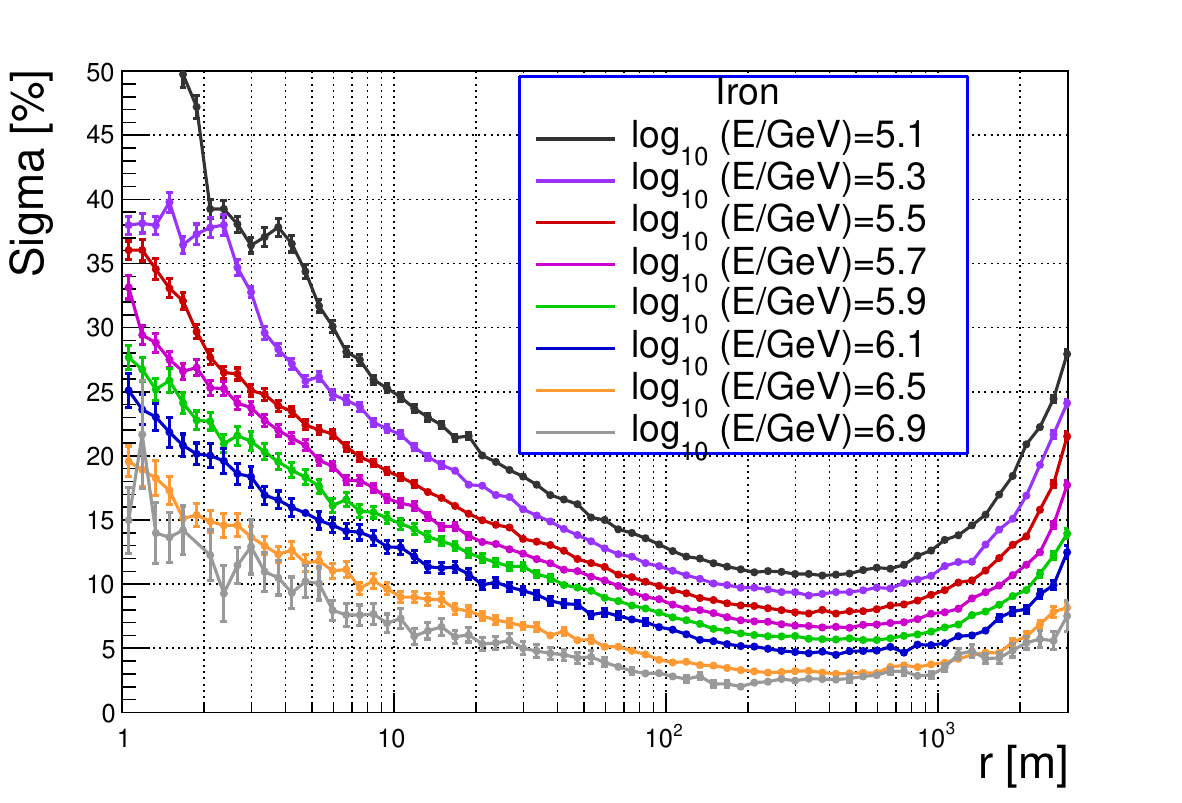}
    \caption{Resolution  in  percentage  (sigma/mean)  of  the particle  number  density  of  secondary  electrons  varies  with perpendicular  distance  to  the  shower  axis.  The  secondary electrons are induced by iron with different energies.
}
    \label{fig:figure9}
\end{figure}

At present, the most commonly used energy reconstruction method is to reconstruct the energy with the secondary particle number density “$\rho$”(electron number density "$\rho_e$", gamma ray number density "$\rho_\gamma$", muon number density "$\rho_\mu$", neutron number density "$\rho_n$") of a certain solid “r” at a given point  \cite{28}. Figure \ref{fig:figure9} shows the variation curve of the expansion percentage of secondary electron density ”$\rho_e$” produced by primary particles of different energies with the position “r” when the primary component is iron. And the lines of different colors represent the different energies of primary particles. It can be seen that for the number density of electrons in the secondary particles, the broadening percentage is smaller in the range of 100-500 m, and it is less dependent on the composition and energy of the primary particles.
Other secondary particles have similar properties, which will not be detailed. Gamma ray at 300-800 m range is better, Muon at 150-600 m range is better and neutron at 800-2000 m range is better.The percentage spread of electron number density at 200 m, gamma ray density at 500 m, Muon density at 250 m, and neutron density at 1000 m will be used to characterize the accuracy of energy reconstruction using them (Figure \ref{fig:figure11}).
\par
\begin{figure}[htbp]
    \centering
    \subfigure
     { \centering
     \label{fig:subfig10_a}
      \includegraphics[scale=0.3]{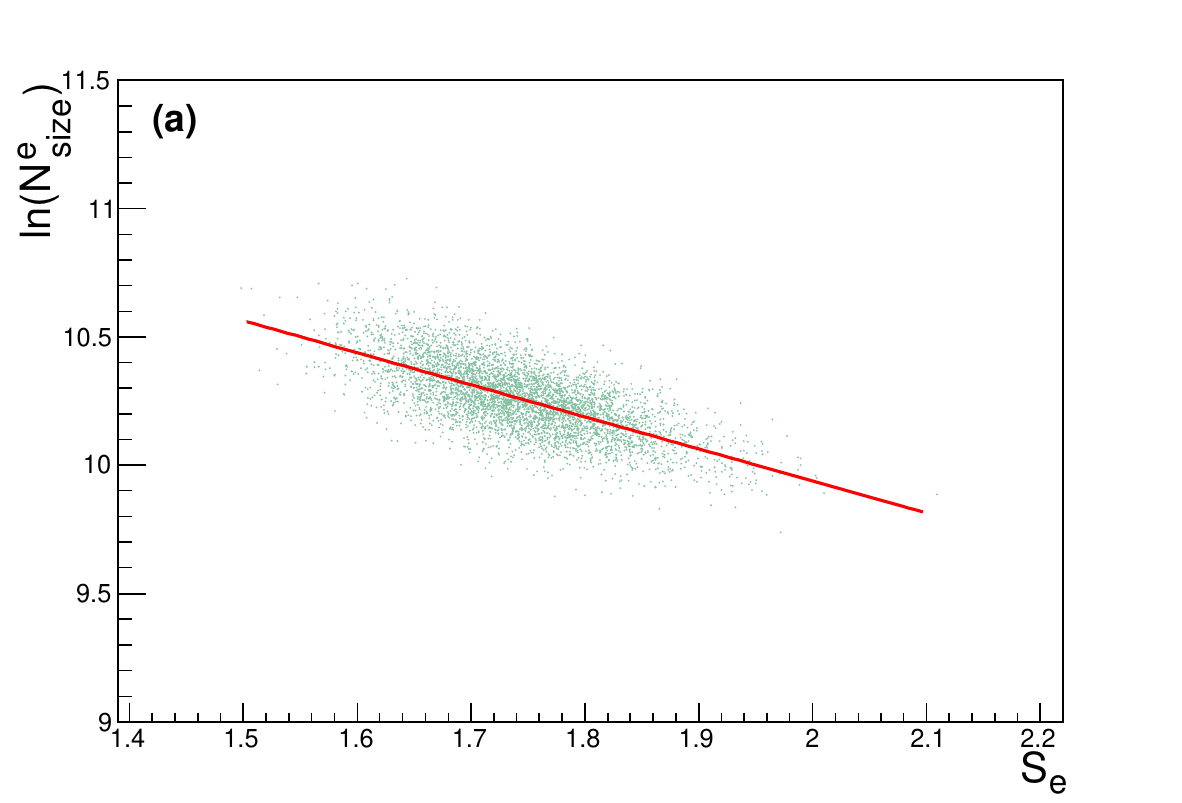}}
    \subfigure
     { \centering
       \label{fig:subfig10_b}
      \includegraphics[scale=0.3]{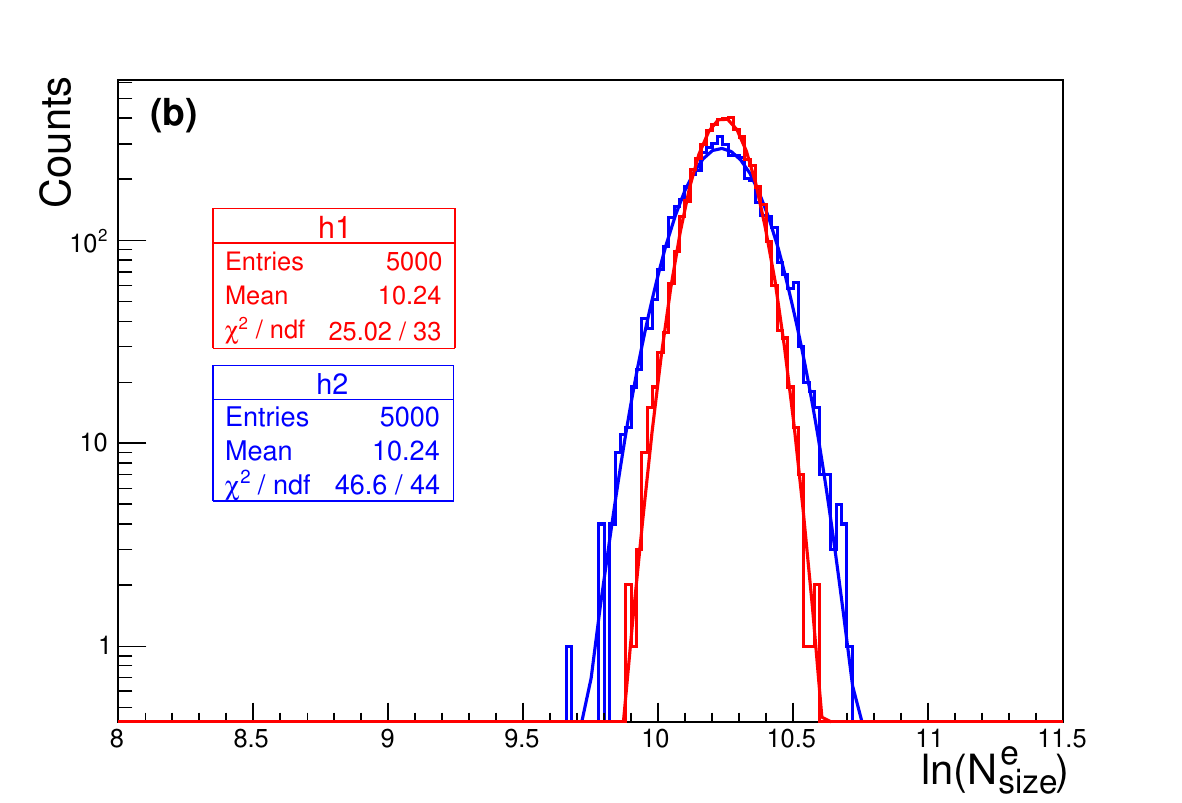}}
    \caption{Distribution of $ln(N_{size}^e)$ vs. $s_e$
from fitted lateral distribution function for iron with energy $log_{10}(E/GeV) = 5.1$ (Red solid
line is a linear fitting); (b) comparison between corrected and uncorrected $\ln \left(N_{\mathrm{size}}^{\mathrm{e}}\right)$, and $\ln \left(N_{\mathrm{size}}^{\mathrm{e}}\right)$ corrected with the red solid line
in panel (a) (In the panel (b), $\chi^
2/ndf$ can represent the good or bad fit, $\chi^2$
represents the difference between the model and the data point, ndf refers to the degree of freedom of fitting, that is the number of data points minus the number of free parameters of the model).}
\label{fig:figure10}
\end{figure}
\begin{figure}[htbp]
    \centering
    \subfigure
      \centering
      \includegraphics[scale=0.3]{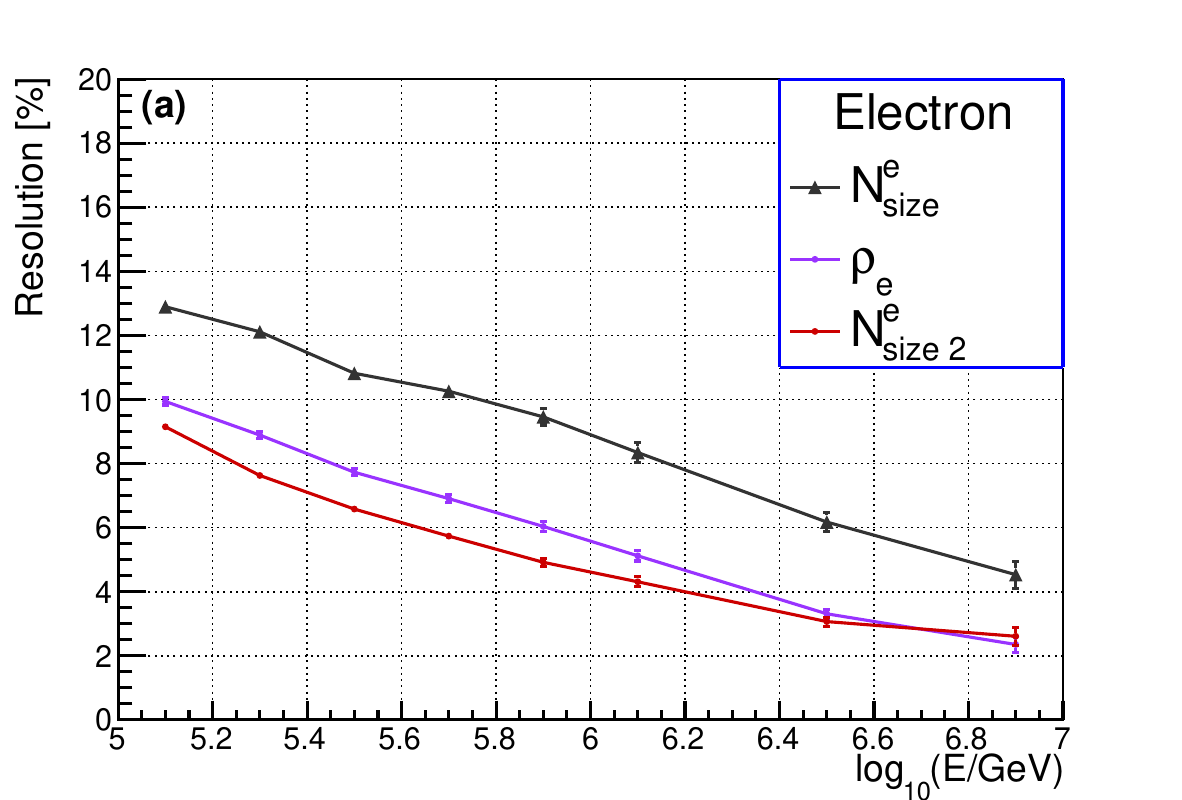}
    \subfigure
      \centering
      \includegraphics[scale=0.3]{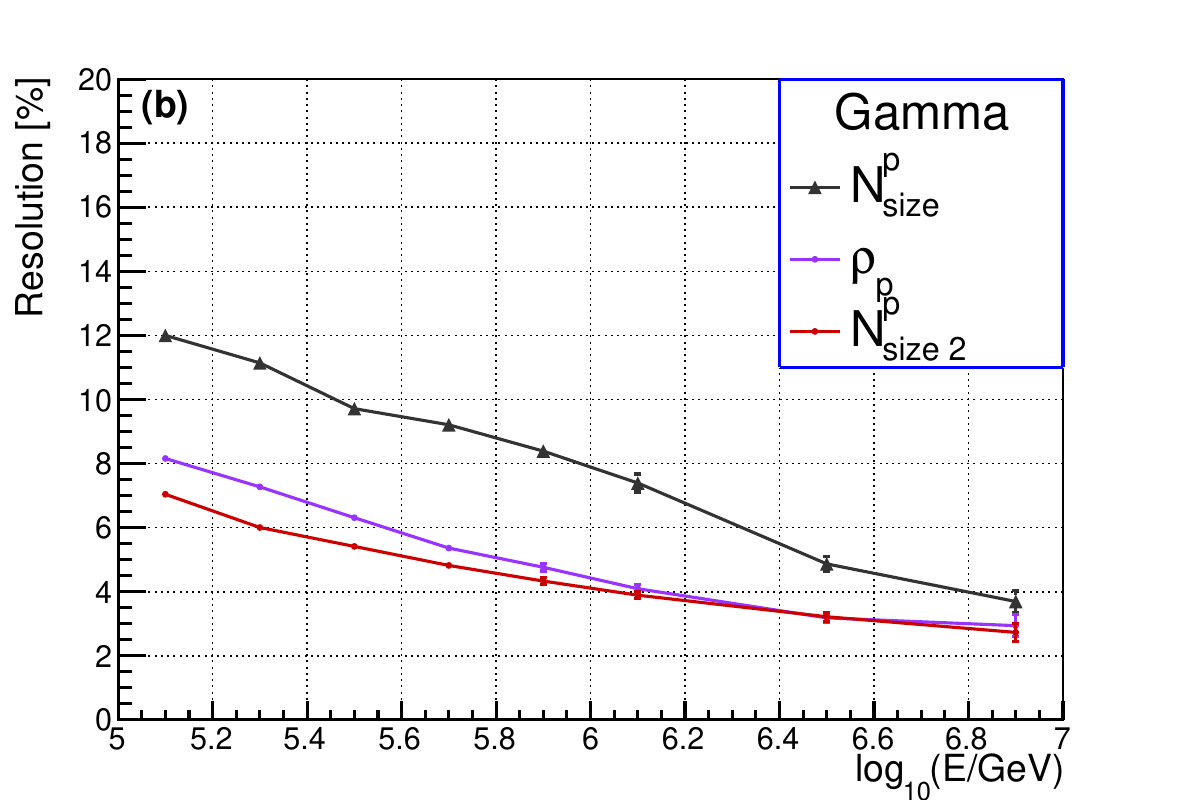}
      \subfigure
      \centering
      \includegraphics[scale=0.3]{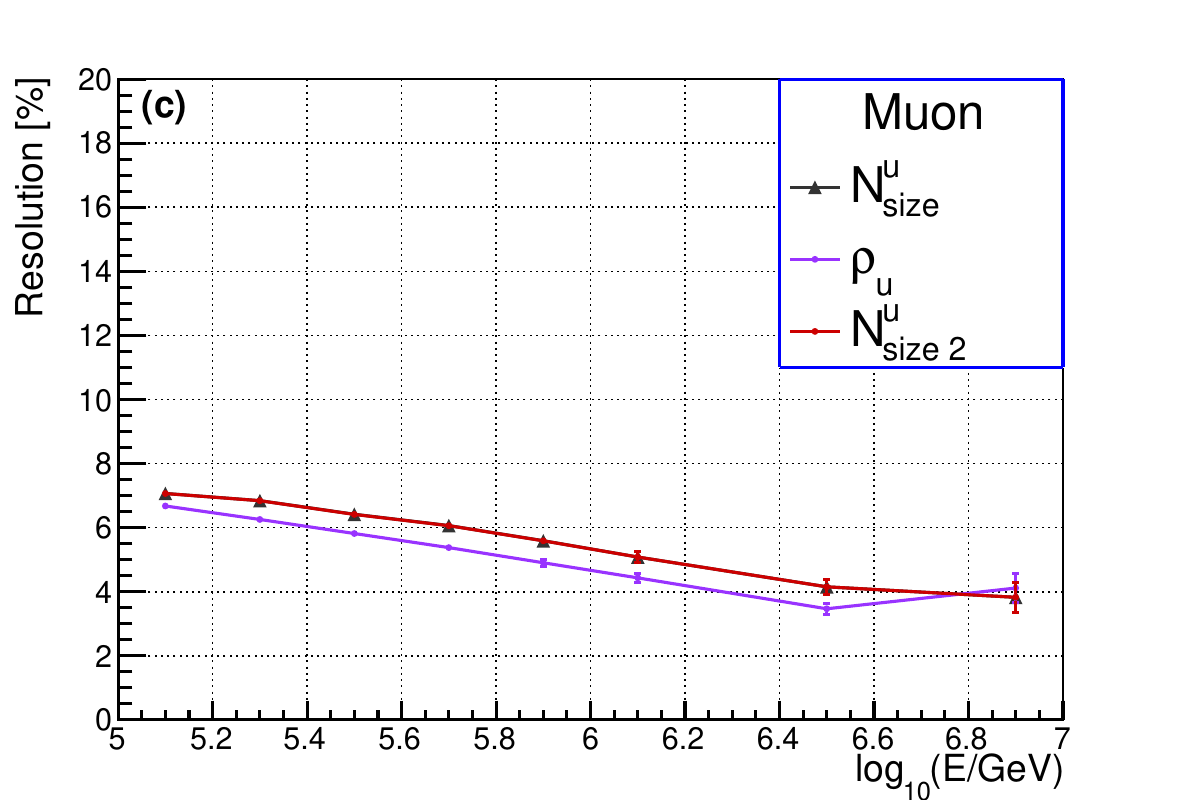}
    \subfigure
      \centering
      \includegraphics[scale=0.3]{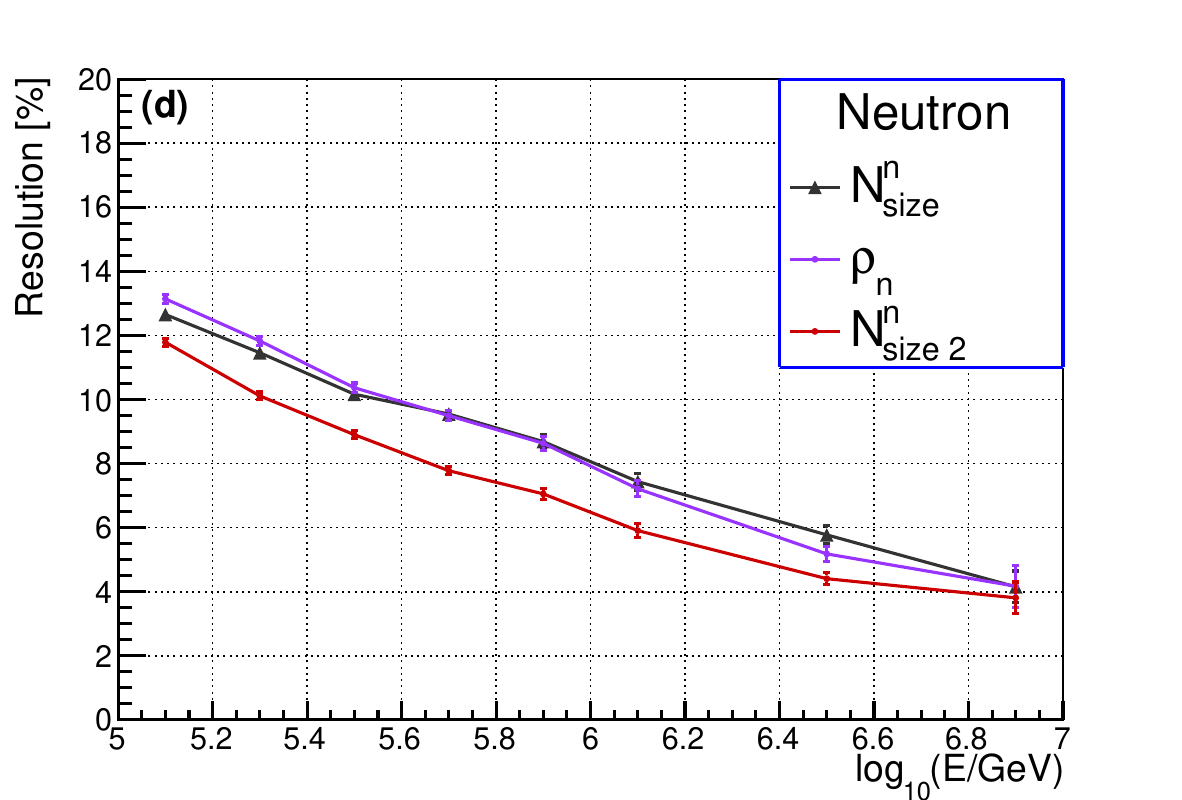}
    \caption{Energy resolution reconstructed by , and respectively. Shower is induced by iron. The secondary particles
are electron (a), gamma (b), muon (c) and neutron (d), respectively;$N_{size}$ indicates the amended $N_{size}$. }
 \label{fig:figure11}
\end{figure}

Another way to reduce the broadening of the particle population
distribution is to use the age parameter “s” to modify the particle
population  \cite{29}. Since the number of secondary particles in 
the observation plane is affected by the development stage of EAS, 
and the age parameter “s” represents the development stage of EAS, 
the influence
of the development stage of EAS can be reduced by modifying the age
parameter. Figure \ref{fig:subfig10_a} shows the distribution of fitting parameters 
of secondary electrons “$N_{\text {size }}$” with “$s_e$” when the original 
particle energy $log_{10}(E/GeV) = 5.1$ and the composition is iron,
which $\ln \left(N_{\mathrm{size}}^{\mathrm{e}}\right)$ decreases 
as se increases. The red solid line is the straight line fitting. The broadening that can be effectively reduced is recorded as” $\ln \left(N_{\text {size } 2}^{\mathrm{e}}\right)$” at the mean corrected to by the red solid line. The red and blue curves in figure \ref{fig:subfig10_b} represent the distribution before and after correction respectively and are fitted with Gaussian functions. The modified broadening of “$\ln \left(N_{\mathrm{size}}^{\mathrm{e}}\right)$” is significantly smaller, which can be used to characterize the energy reconstruction accuracy of this method and the results are shown in figure \ref{fig:figure11}. Figure \ref{fig:figure11} shows the comparison of the energy reconstruction accuracy before and after age correction  “$N_{\text {size }}$” “$N_{\text {size } 2}$” and the secondary particle population density for energy reconstruction when the primary particle was iron, as well as the relationship between the energy reconstruction accuracy and the primary particle energy. It can be seen that for electrons and gamma rays in secondary particles, the energy reconstruction accuracy can be effectively improved by using the particle number and particle number density modified by age parameter compared with the direct use of particle number, and the energy reconstruction accuracy of the particle number modified by age parameter 
is slightly better than that obtained by particle number density. But the difference is not large. For muons and neutrons in secondary particles, there is little difference in the energy reconstruction accuracy of the particle number, particle density and particle number modified by age parameter. Since the modified curve of the age parameter depends on the energy and type of the original particle, and the radius selected for the particle number density is a fixed value, and the energy reconstruction accuracy obtained between them is not very different,
the particle number density of the secondary particle at a fixed point will be used to characterize the optimal energy reconstruction accuracy. The results here are similar for the energy reconstruction accuracy of the EPOS-LHC model and the QGSJet-Ⅱ-04 model. In this paper, the limit detection
performance under ideal conditions is studied and the process of detector response is not involved. The systematic error of energy 
reconstruction caused by the difference of the mean particle 
numbers of the two strong interaction models is not considered here. 

\begin{figure}[htbp]
    \centering
    \includegraphics[width=0.5\textwidth]{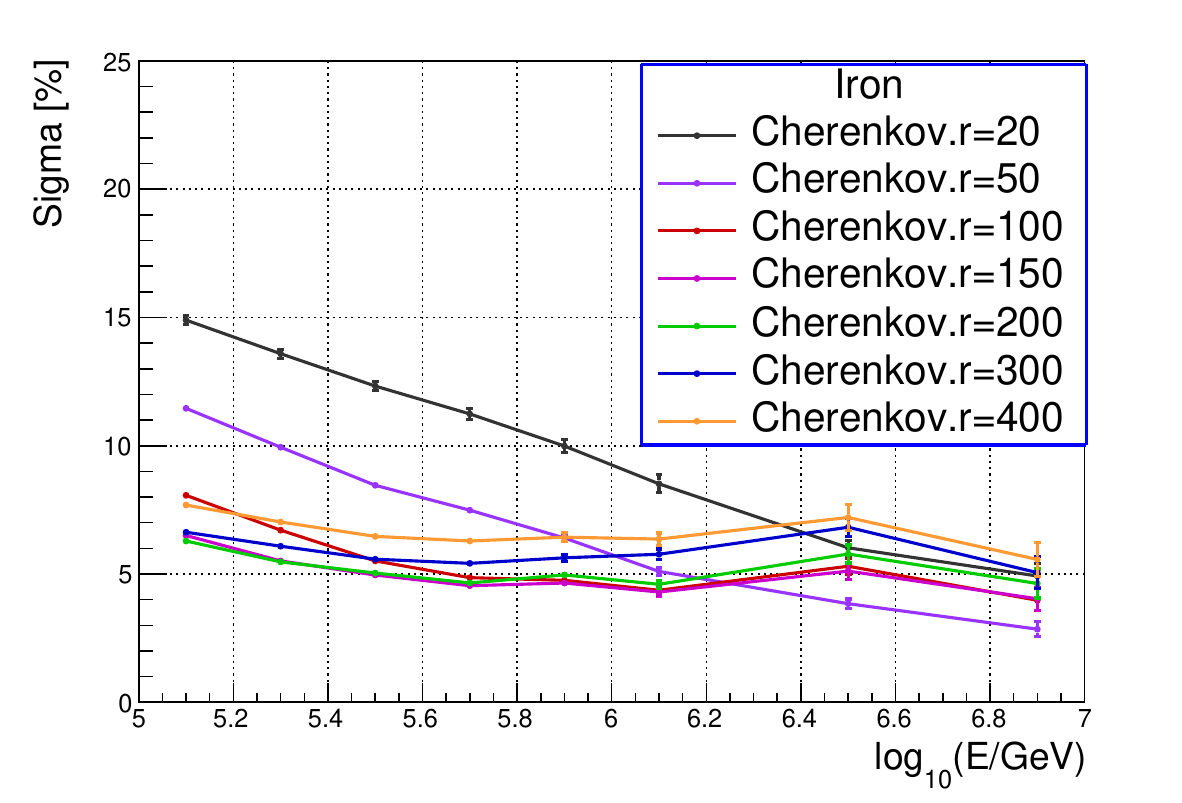}
    \caption{ Resolution in  percentage  (sigma/mean)  of  Cheren kov  photon  number $N^C$ varies  with  the  energy  of  the primary  particle  at  different  vertical  distance r from  the core site. Shower is induced by iron.
}
    \label{fig:figure12}
\end{figure}

Figure \ref{fig:figure12} shows the percentage broadening of the distribution of Cherenkov photon numbers ($N^C$) at different vertical distances from the core site when the primary particle is iron, and the vertical distances from the center to the shower axis are 20, 50, 100, 150, 200, 300 and 400 m, respectively. The number of Cherenkov photons at visible of 50m, 150m and 200 m has a smaller spread of about 4\%-7\%. 
\par
\begin{figure}[htbp]
    \centering
    \subfigure
     { \centering
     \label{fig:subfig13_a}
      \includegraphics[scale=0.2]{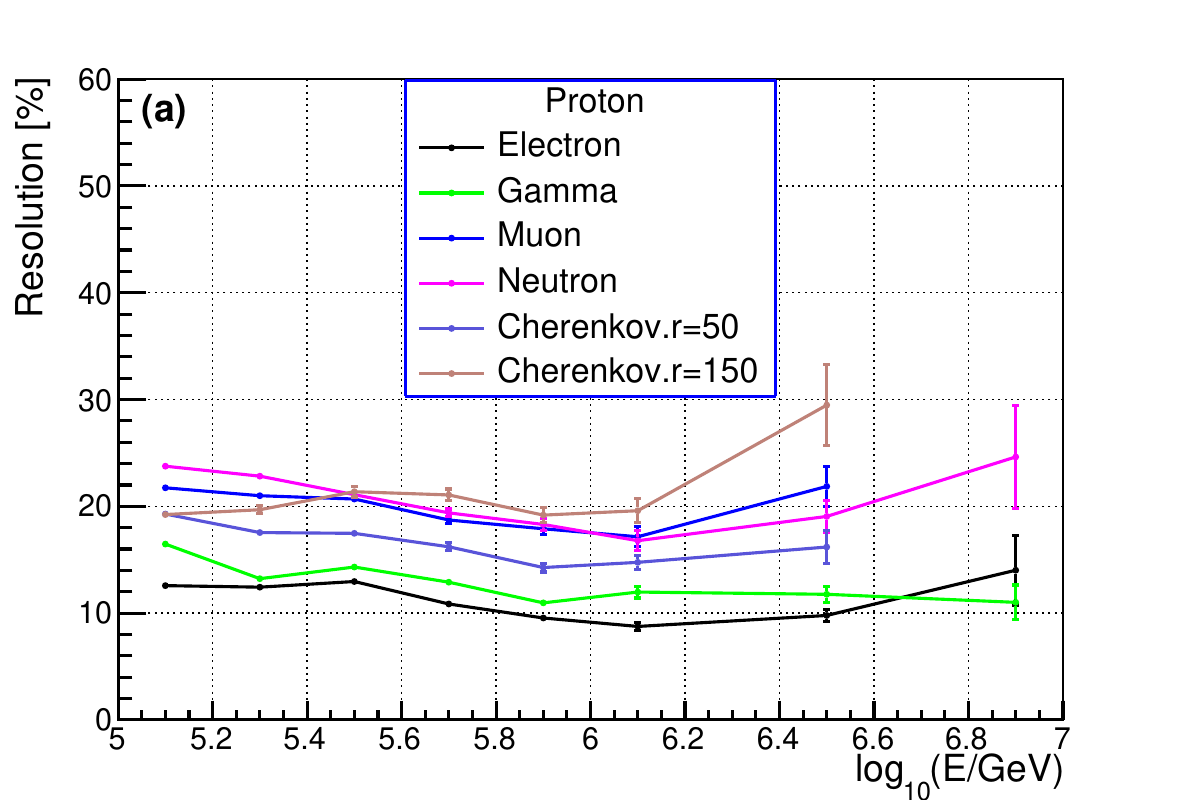}}
   \subfigure
     { \centering
  \label{fig:subfig13_b}
       \includegraphics[scale=0.2]{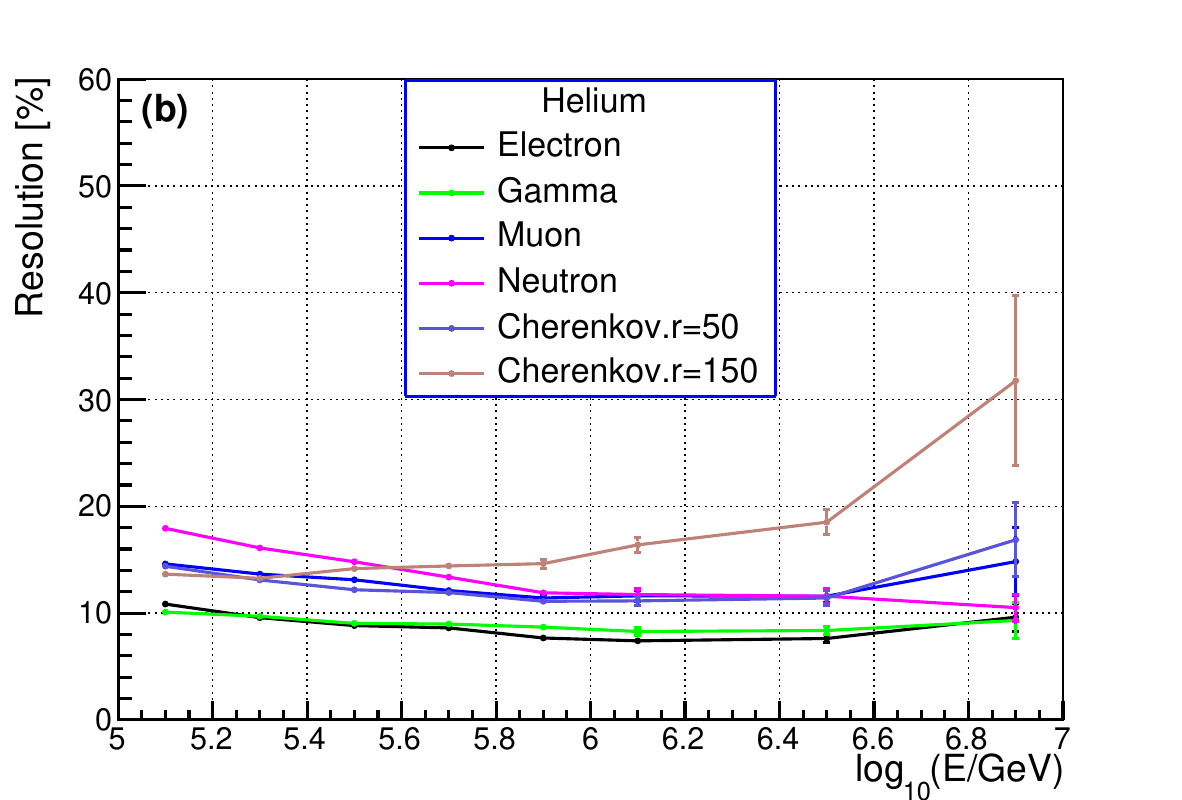}}
      \subfigure
      { \centering
       \label{fig:subfig13_c}
      \includegraphics[scale=0.2]{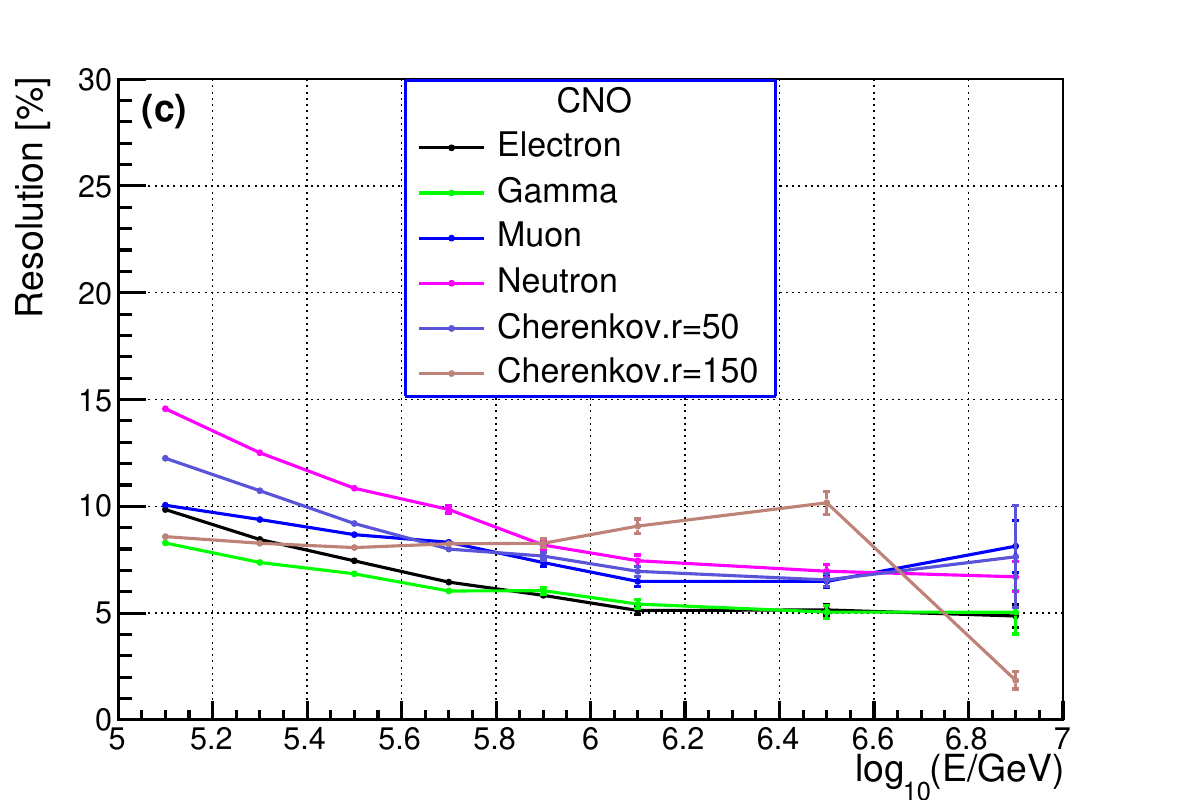}}
     \subfigure
     { \centering
    \label{fig:subfig13_d}
      \includegraphics[scale=0.2]{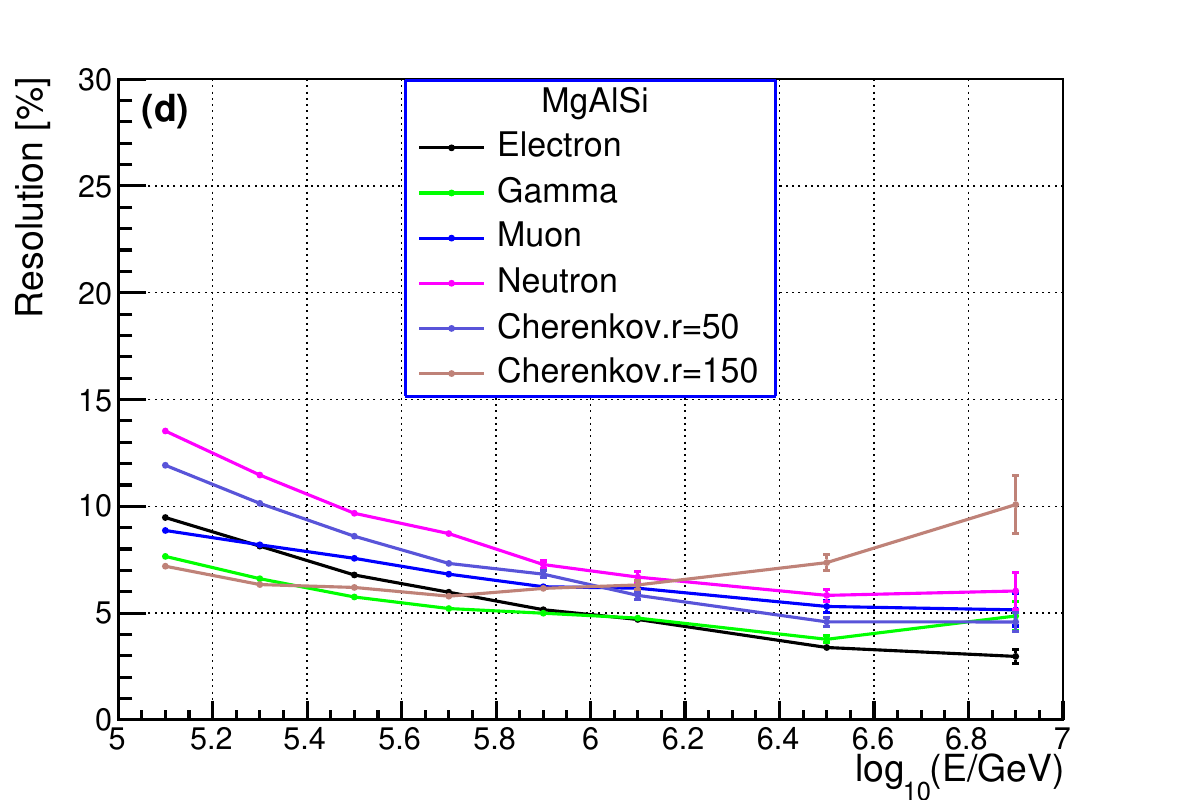}}
      \subfigure
      { \centering
     \ \label{fig:subfig13_e}
       \includegraphics[scale=0.2]{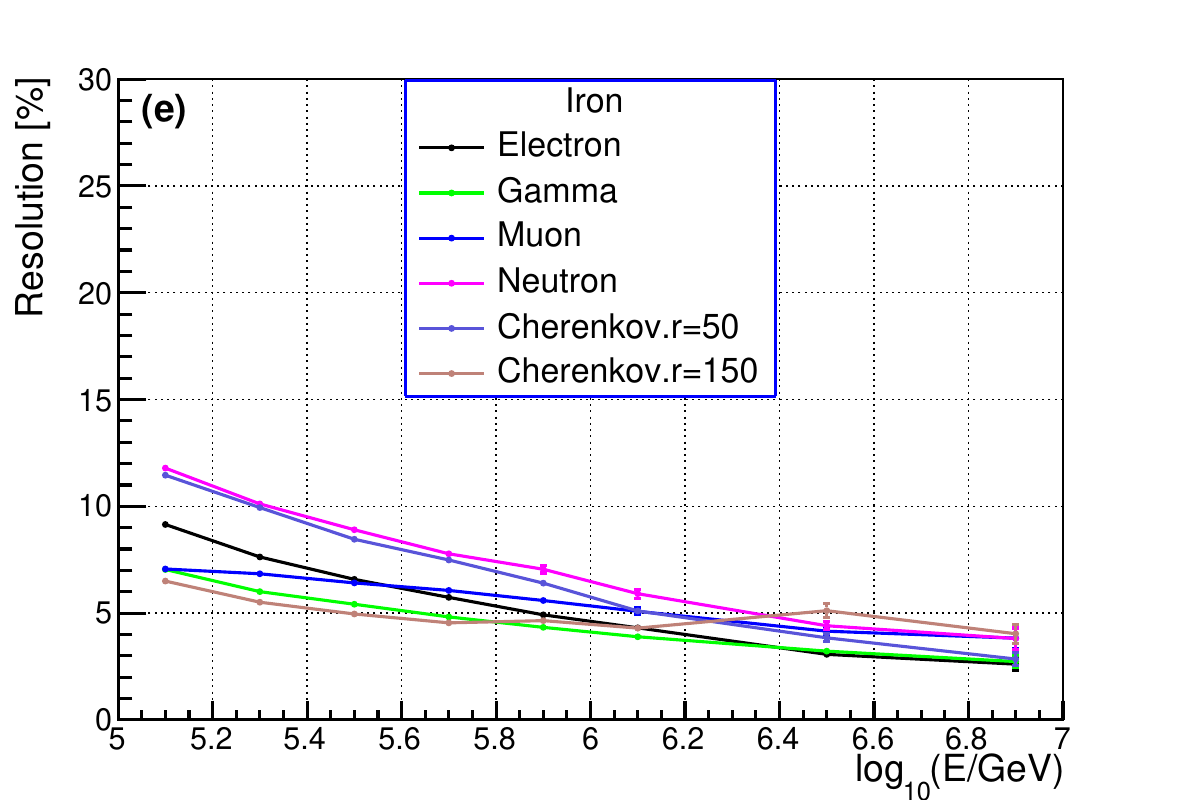}}
    
    \caption{Energy  resolution  from  different  secondary  components  vs.  primary  particle  energy.  The  primary  particles  are  proton  (a), helium (b), CNO (c), MgAlSi (d), iron (d) (Colored lines indicate different secondary type)}
 \label{fig:figure13}
\end{figure}

The energy reconstruction accuracy obtained by different secondary components is shown in figure \ref{fig:figure13}, where the primary particles in Figure 13(a) - (e) correspond to proton, helium, carbon, nitrogen, oxygen, magnesium, aluminum, silicon and iron, respectively. For protons, the energy reconstruction accuracy of electrons, gamma rays and Cerenkov photons at 50 m is about 10\%-19\%. For iron, the energy reconstruction accuracy of gamma rays, Cherenkov photons and muons at 150 m is better and is about 4\%-8\%. The higher the mass number of the primary particle, the higher the precision of energy reconstruction. In the experiment, multiple secondary particles can be combined according to the accuracy of individual energy reconstruction of different secondary particles to obtain energy reconstruction variables with less component dependence and higher precision. The above results can provide references for the selection of secondary particle types, energy reconstruction methods and distance from the core.

\section{Composition discrimination}
\label{sec:5}
The identification of the primary particles of cosmic rays is the key to the single component energy spectrum measurement of cosmic rays. The study on the primary particle sensitivity of secondary components in EAS can provide guidance for the selection of component identification variables. In this paper, the ability of these parameters to identify primary particles is studied according to the fitting parameters of the transverse distribution of secondary particles obtained in Section \ref{sec:3}. Since Cherenkov light is mainly used to identify primary particles based on imaging, it is beyond the scope of this study and will not be studied. According to the study in Section \ref{sec:4}, it can be seen that the particle number density fluctuates less than the total number of particles, and it will also have better discrimination ability for component identification. Here, the particle number density will be used instead of the total number of particles to study the particle discrimination ability. 
\par
This paper will study the component identification ability of each variable under the same energy condition, which can show the independent identification ability of each variable. For real experimental data, energy reconstruction variables can be combined with energy-independent component identification variables. For example, the number of electromagnetic particles and the number of muse of secondary particles are related to the energy and type of the original particle. The fluctuation of electromagnetic particles is smaller and the number of muon is more sensitive to the composition. So it can be modified on the basis of electromagnetic particles and the number of muon can be used to obtain the energy reconstruction variable independent of the composition. It is modified using the previous energy reconstruction variable to obtain an energy-independent component sensitive variable. For reasons of space, I will not go into details here. 
\par
\begin{figure}[htbp]
    \centering
    \includegraphics[width=0.5\textwidth]{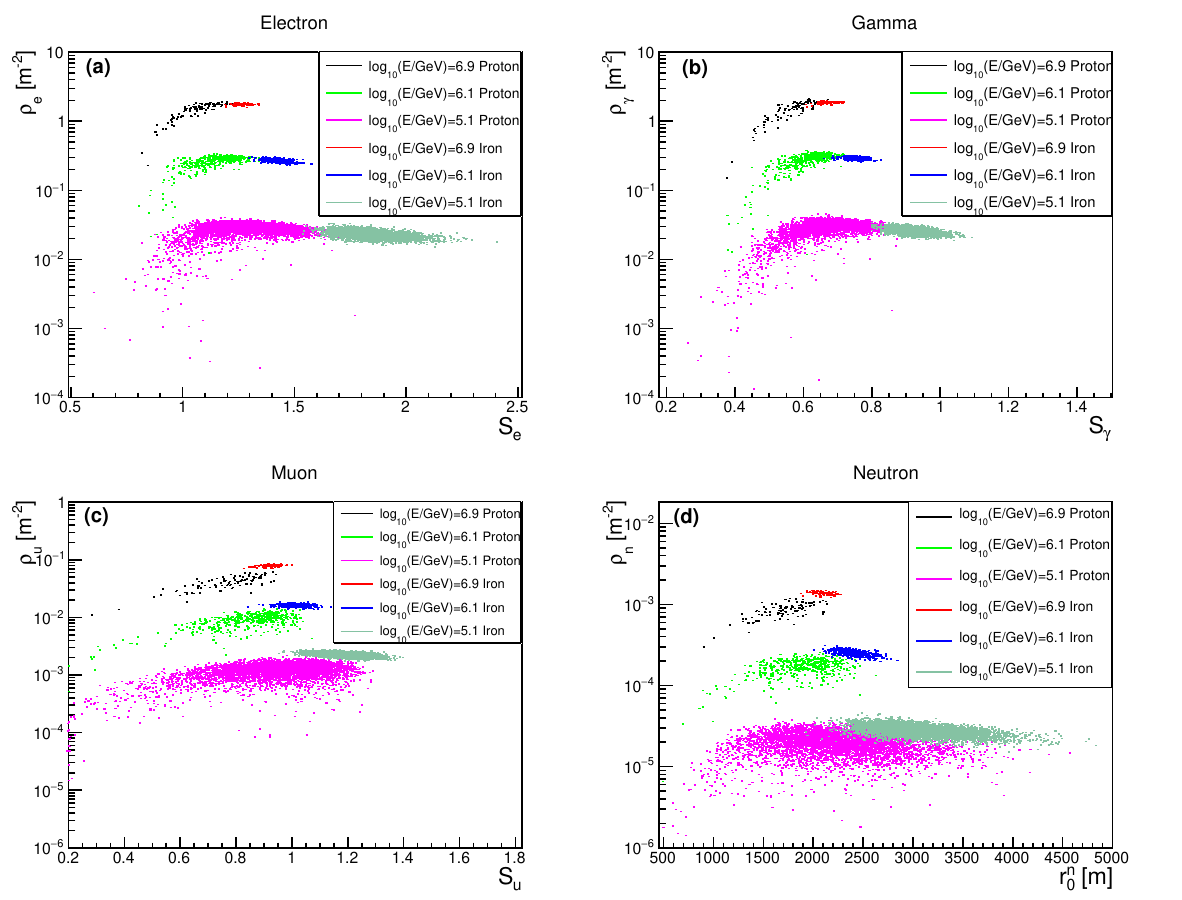}
    \caption{Distribution of r vs. s when shower is induced by proton and iron respectively with different energies: (a)$\rho_e$ vs.$s_e$ ;(b)$\rho_\gamma$  vs.  $s_\gamma$; (c)$\rho_\mu$ vs. $s_\mu$ ; (d)$\rho_n$ vs. $r_0^n$.}
    \label{fig:figure14}
\end{figure}
\begin{figure}[htbp]
    \centering
    \includegraphics[width=0.5\textwidth]{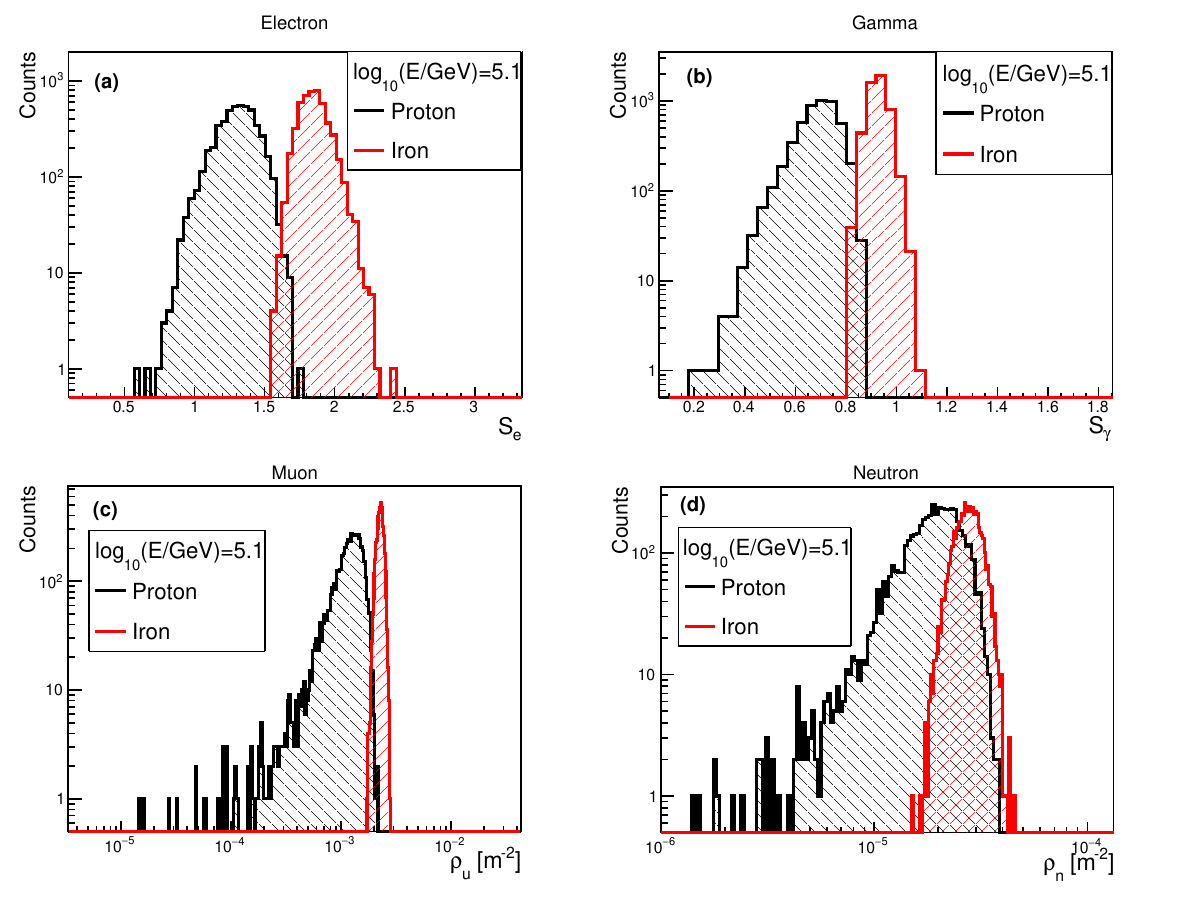}
    \caption{Comparison of the distribution of (a) $s_e$ , (b) $s_\gamma$ , (c)
$\rho_\mu$ , (d) $\rho_n$ between proton and iron with $log_{10}(E/GeV) = 5.1$}
    \label{fig:figure15}
\end{figure}
\begin{figure}[htbp]
    \centering
    \includegraphics[width=0.5\textwidth]{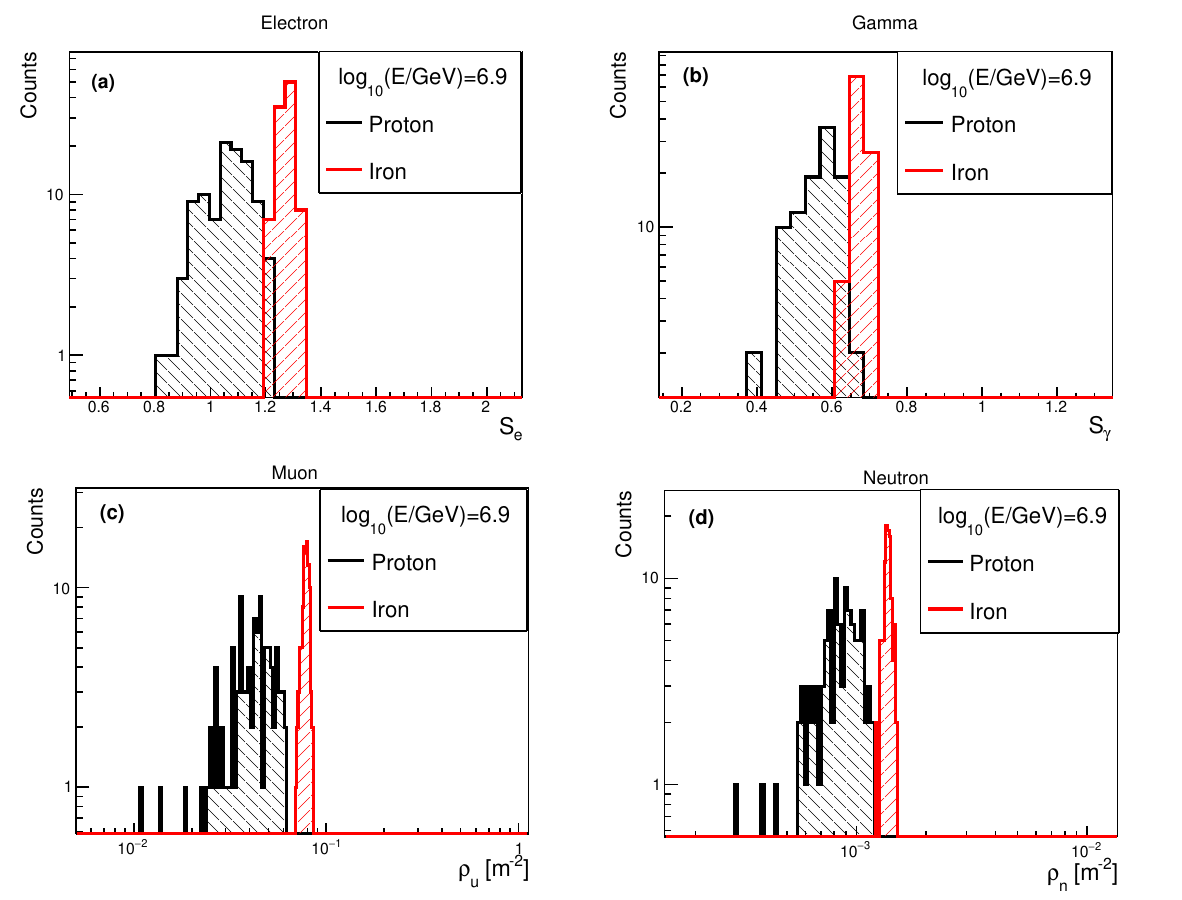}
    \caption{Comparison of the distribution of (a) $s_e$ , (b) $s_\gamma$ , (c)
$\rho_\mu$ , (d)
$\rho_n$ between proton and iron with $log_{10}(E/GeV) = 6.9$}
    \label{fig:figure16}
\end{figure}

Figure \ref{fig:figure14} shows the ability of particle number density and age parameters to distinguish protons and iron nuclei when the secondary particles are positrons, gamma rays, muons and neutrons respectively. The dots of different colors in Figure \ref{fig:figure14} represent protons and iron nuclei with energy $log_{10}(E/GeV) = 5.1$, $log_{10}(E/GeV) =6.1$ and $log_{10}(E/GeV) = 6.9$, respectively. It can be seen that the age parameters “s” of positrons and gamma-rays, the particle number density of muons and neutrons have a good ability to identify the composition. In order to more vividly demonstrate their particle identification ability, the distributions of protons and iron nuclei at the same energy are projected onto the coordinate axes of the above variables, respectively, as shown in figure \ref{fig:figure15} and figure \ref{fig:figure16}. It can be seen that the identification energy of muon particle number density is the best in both low energy and high energy segments. The age parameter “$s_e$” ”$s_\Gamma$” of positron and gamma ray transverse distribution shape is better in low energy segment (such as around 100 TeV) and the identification ability of neutron particle number density is better in high energy segment (such as around 10 PeV). In the experiment, several secondary particles can be combined to obtain the component identification variable with less energy dependence and better identification ability according to the identification ability of different secondary particles. This can provide reference for the selection of component identification variables and detector types at different energies.

\par
\begin{figure}[htbp]
    \centering
    \subfigure
      \centering
      \includegraphics[scale=0.2]{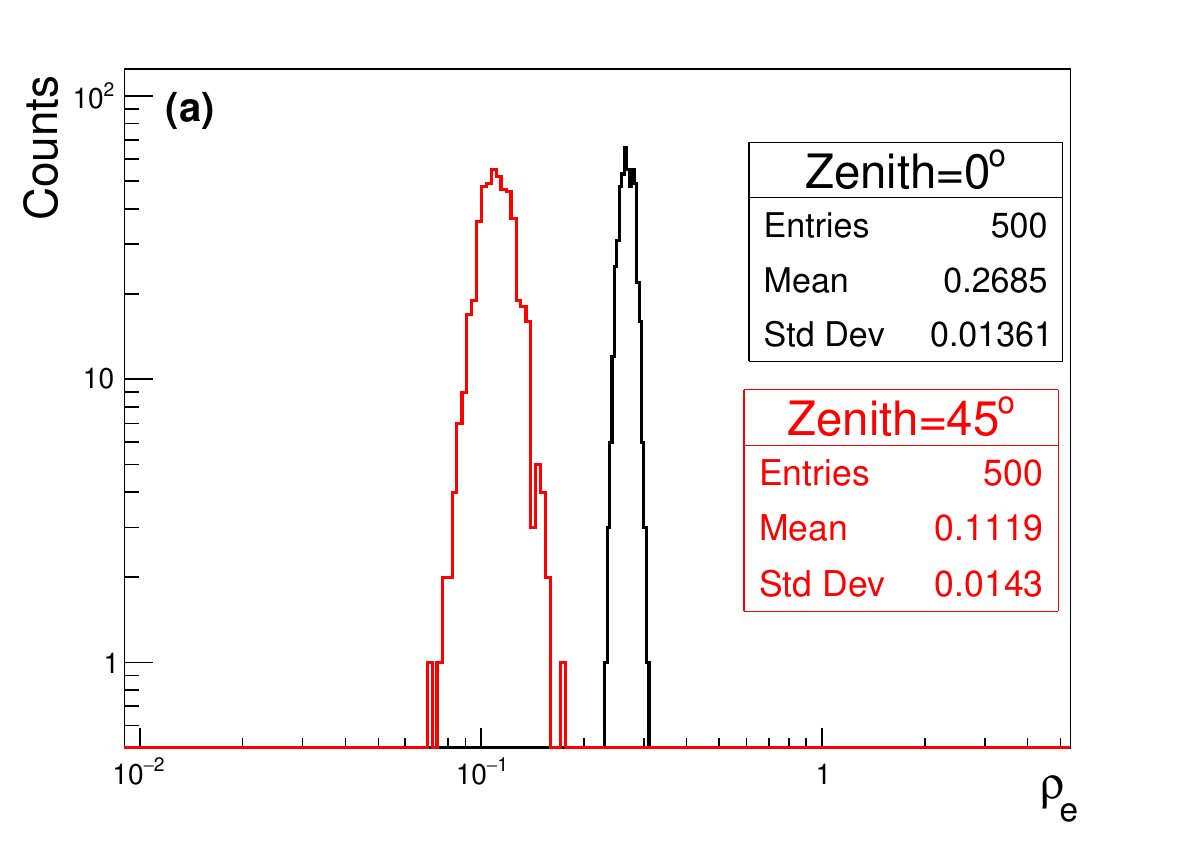}
    \subfigure
      \centering
      \includegraphics[scale=0.2]{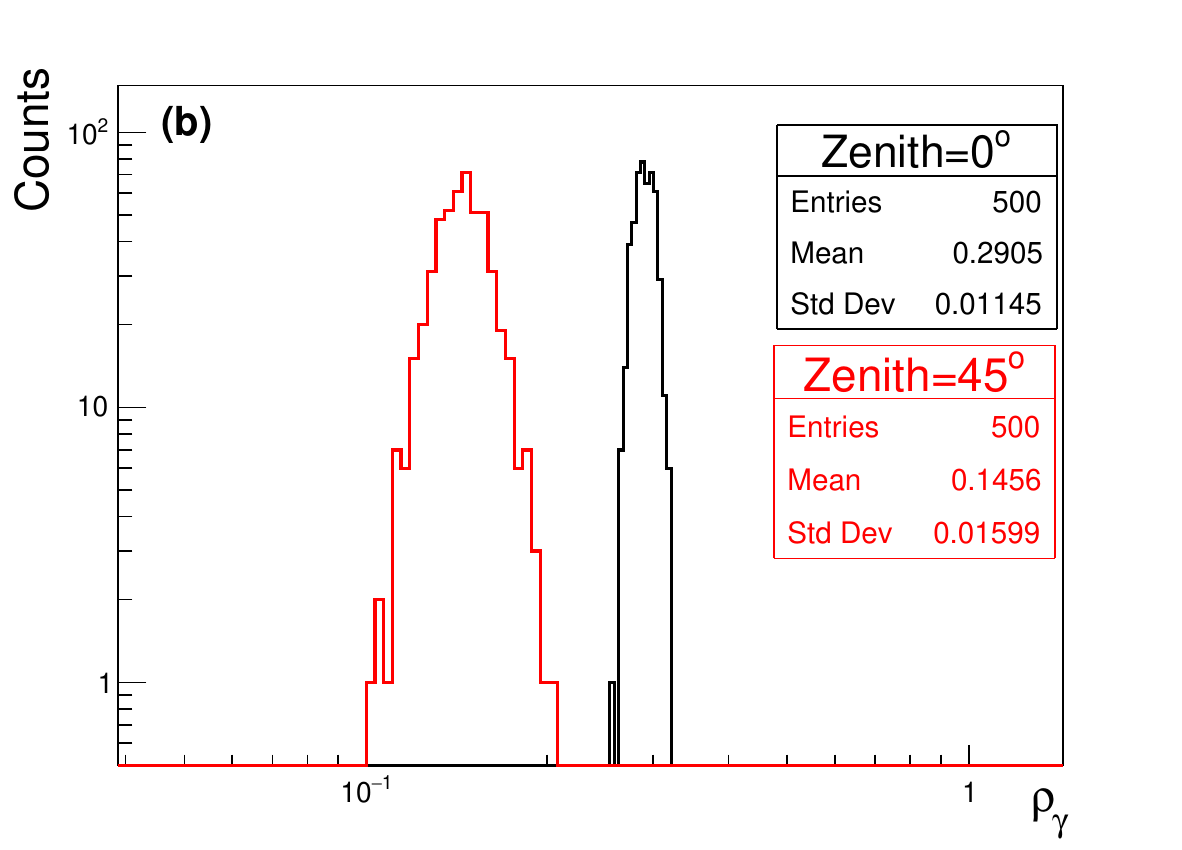}
    \subfigure
      \centering
      \includegraphics[scale=0.2]{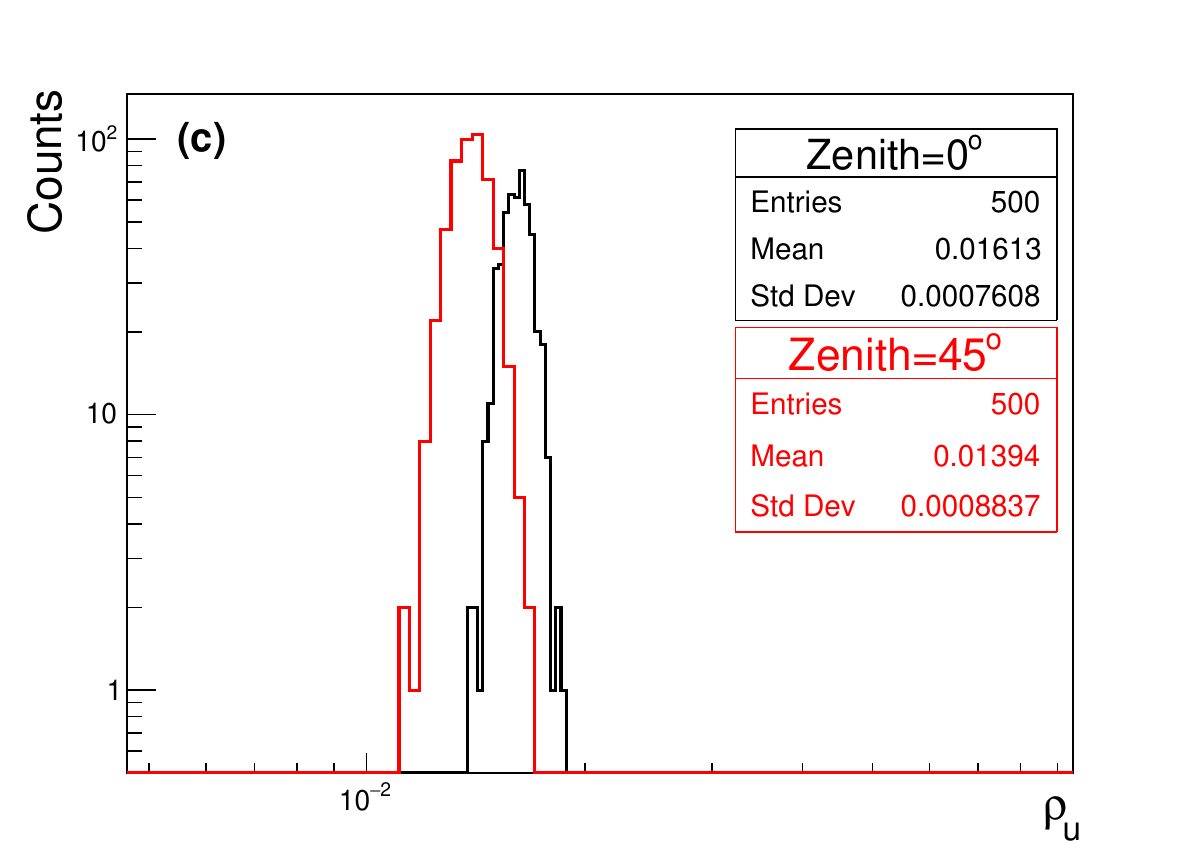}
    \subfigure
      \centering
      \includegraphics[scale=0.2]{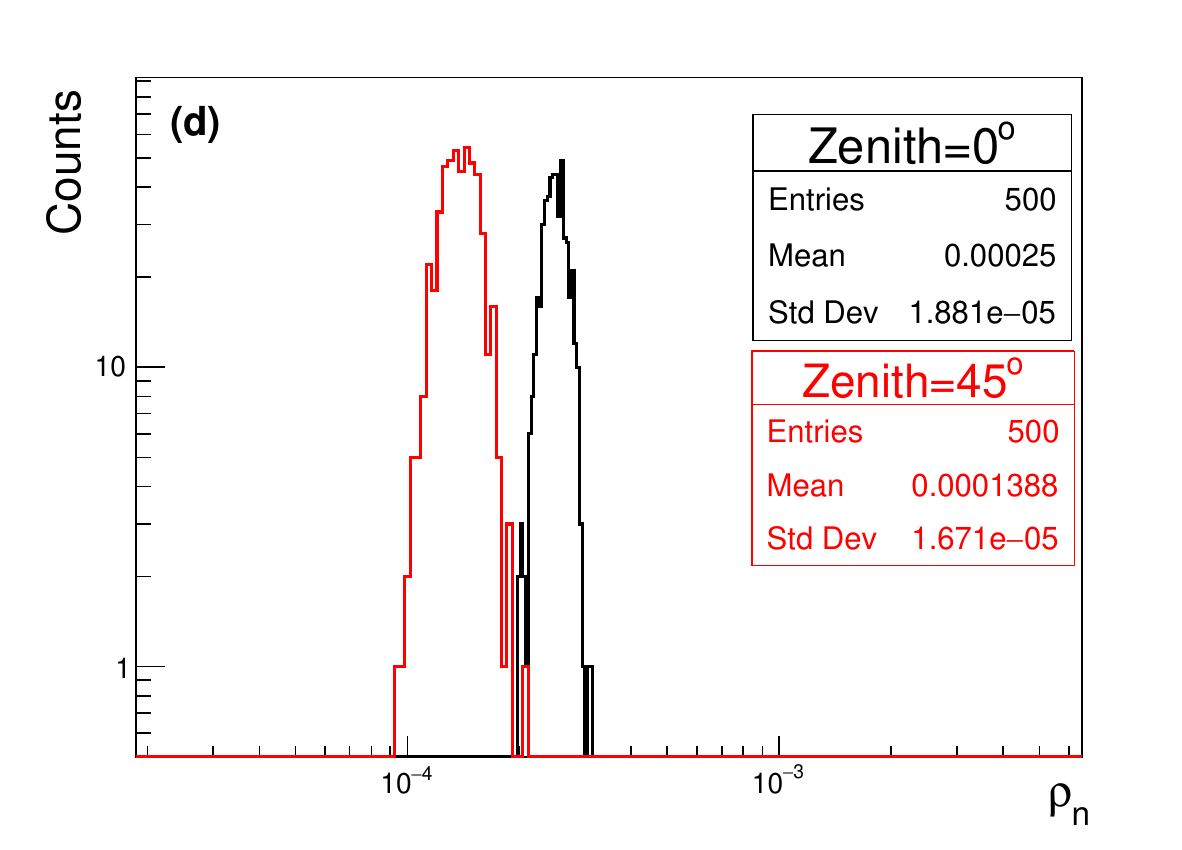}
    \subfigure
      \centering
      \includegraphics[scale=0.2]{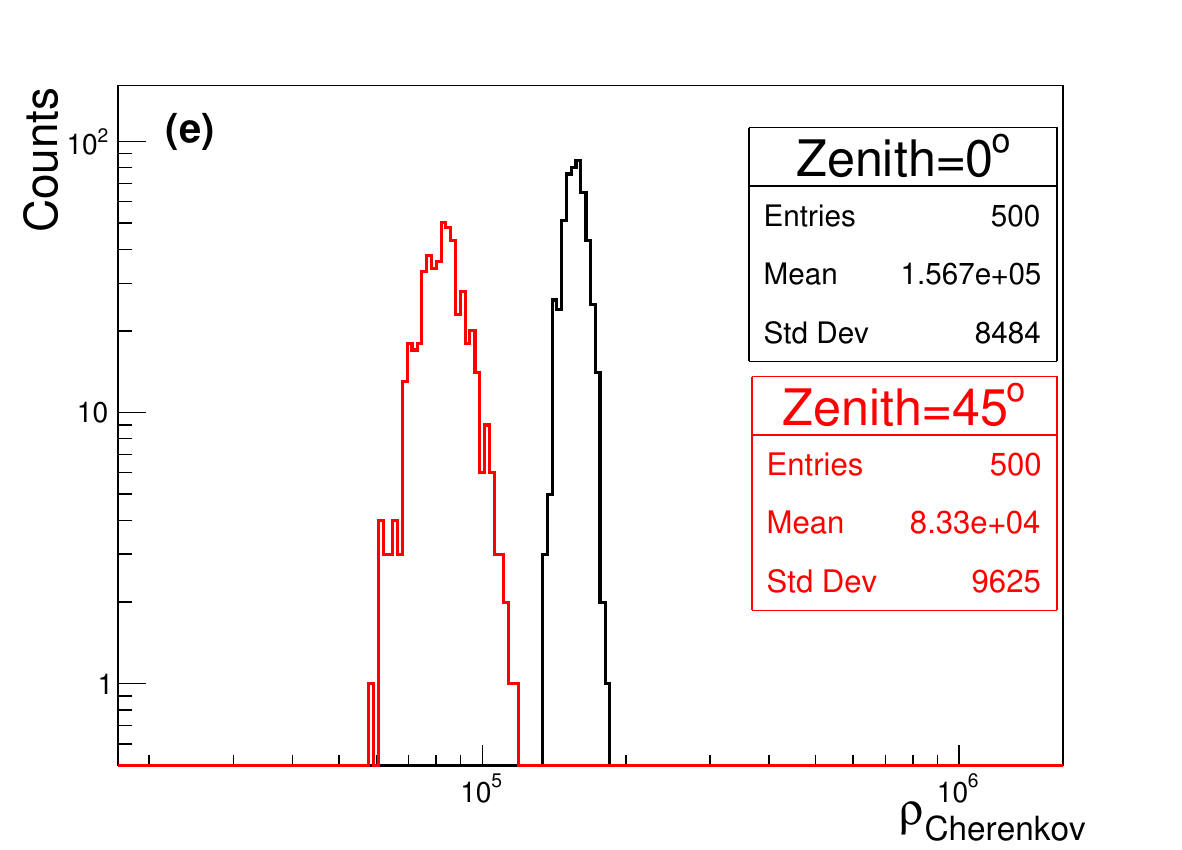}
      \subfigure
      \centering
      \includegraphics[scale=0.2]{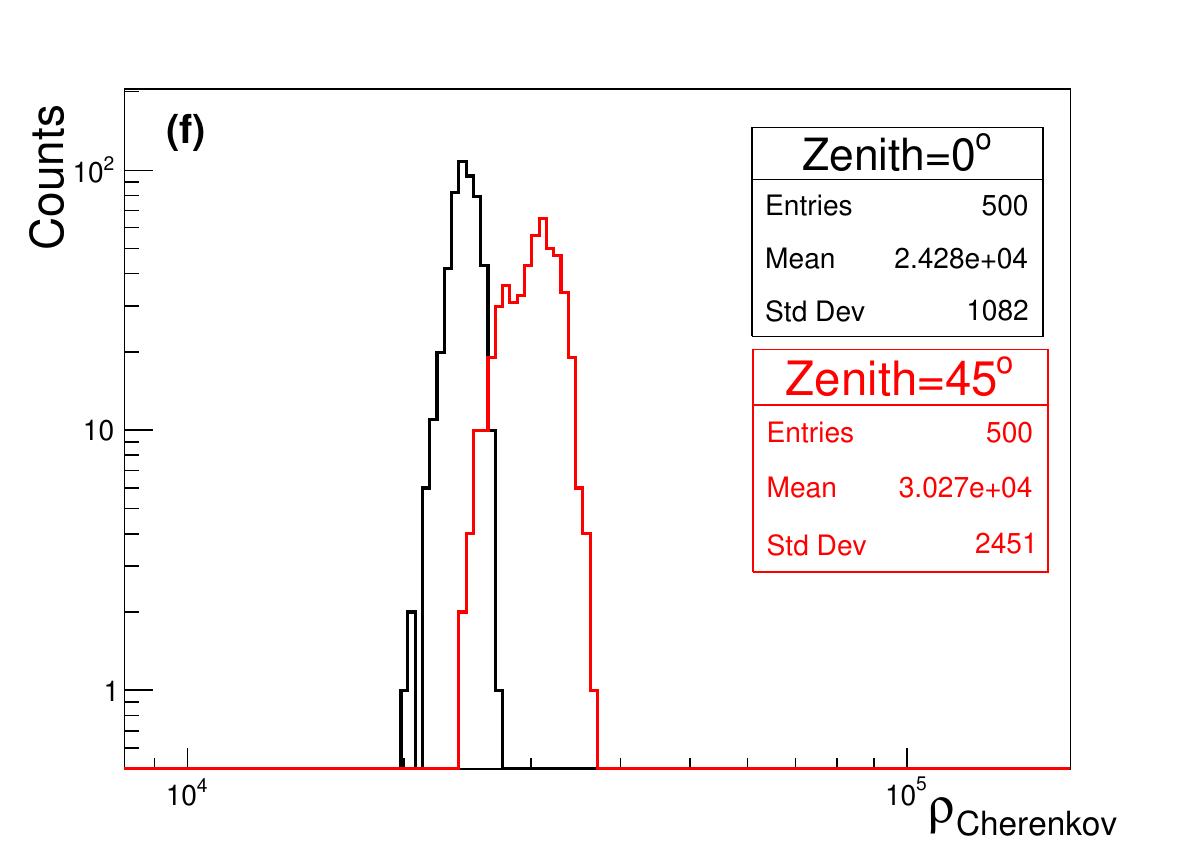}
     
    \caption{Comparison of the distribution of
$\rho_e$ (a), $s_\gamma$ (b),
$\rho_\mu$ (c),
$\rho_n$ (d) and
$\rho_C$ at r=50 m (e), r=150 m (f) for $\theta=0^{\circ}$ and $\theta =45^{\circ}$.
The primary particle is iron with energy $log_{10}(E/GeV)=6.1$.}
\label{fig:figure17}
\end{figure}

This paper also studies the case of a zenith Angle of 45°, which will increase the atmospheric depth compared to the case of vertical incidence. For the energy segment studied in this paper and the selected altitude, the zenith angle is the atmospheric depth of 45° will exceed the atmospheric depth required for the shower to develop to the maximum area, while the atmospheric depth of vertical incidence is near the shower to develop to the maximum area. So the distribution of detected secondary components will be affected and the energy reconstruction accuracy and particle identification ability will be affected. The distribution comparison of secondary components at different zenith angles is shown in figure \ref{fig:figure17}. It can be seen that as the zenith Angle increases, the number of electrons and gamma rays will decrease because they have passed through the shower maximum and their fluctuations will also be significantly larger. The muse interact less with the atmosphere during propagation, the increase of atmospheric depth has little influence on the muse number and the fluctuation of the muse number is also small. Neutrons also continue to undergo hadron shower processes with the atmosphere, attenuating more relative to the muon and the fluctuation of neutrons increases with the increase of the zenith, the amplitude is between the electromagnetic particles and the muon. The telescope measures the Cherenkov light at a fixed position, and the density of the Cherenkov light varies with the zenith angle depending on the vertical distance of the detection area from the shower axis ”r”. As shown in figure \ref{fig:figure17}, when r= 50 m, the number of Cherenkov photons decreases with the increase of zenith angle. While, the number of Cerenkov photons increases with the increase of zenith Angle at r= 150 m. At r= 50 m and r= 150 m, the fluctuation of Cherenkov photon number increases significantly and the variation amplitude is similar to the fluctuation amplitude of electromagnetic particles. In summary, when the atmospheric depth exceeds the atmospheric depth where the shower develops to the maximum, the fluctuation of various secondary components increases, the fluctuation of Muse changes the least and the fluctuation of electromagnetic particles changes the most.
\section{Summary}
\label{sec:6}
The single-component energy spectrum of the cosmic ray knee region is an important means to understand the physical origin of the cosmic ray knee region. Since the ground experiment lacks a good absolute energy calibration method, and can only measure secondary particles produced by the primary particles in EAS and cannot directly measure the primary particles. The energy measurement and particle identification ability of the ground experiment are called the limiting factors of single-component energy spectrum measurement. Based on this, we simulate the characteristics of secondary components of cosmic rays with different energies and primary particle compositions at 4400 m above sea level in EAS, including positrons, gamma rays, muons, neutrons and Cherenkov photons. The lateral distribution characteristics of various secondary components and the strong interaction dependence of the lateral distribution of different secondary components are studied in detail and the lateral distribution is fitted well with specific functions. The method and accuracy of energy reconstruction, strong interaction model dependence and discriminability of cosmic-ray particle identification are studied in detail with these fitting parameters. It provides reference for the selection of detector types, energy reconstruction methods and component identification variables in various ground experiments.
\par
For energy reconstruction, using the number density of secondary particles at a certain vertical distance from the core site ”r” is a better choice than the total number of secondary particles. It is less compositional dependent than the total number of particles modified by the age parameter. When the primary particle is proton, the energy reconstruction accuracy of electron, gamma ray and Cherenkov photon at 50 m is good and is about 10\%-19\%. When the primary particle is iron, the energy reconstruction accuracy of gamma ray, Cherenkov photon and Muon at 150 m is better and is about 4\%-8\%. The higher the mass number of the primary particle, the higher the precision of energy reconstruction. In the experiment, multiple secondary particles can be combined according to the accuracy of individual energy reconstruction of different secondary particles to obtain energy reconstruction variables with less component dependence and higher precision.
\par
For particle identification, the identification energy of muon particle number density "$\rho_\mu$" is the best in both the low energy segment and the high energy segment. The age parameter of positron and gamma ray transverse distribution shape is better in the low energy segment (such as around 100 TeV) and the identification ability of neutron particle number density is better in the high energy segment (such as around 10 PeV). In the experiment, multiple secondary particles can be combined to obtain component identification variables with lower energy dependence and better identification ability according to the identification ability of different secondary particles, such as multivariate analysis method  \cite{30} or deep learning method \cite{31}. The parameters provided in this paper can be directly used as training variables.
\par
For the difference in transverse distribution of secondary particles caused by the two strong interaction models of EPOS-LHC and QGSJET-II-04, the difference in the number of positrons, gamma rays and Cherenkov photons is very close and they are smaller than that of muon and neutrons and it is within 5\% at r$>$20m and within 10\% at all ranges of r. The difference in the number of muon is within 5\% at r$>$100m. But the maximum difference can be close to 20\% (corresponding to the original particle is iron, the energy is about 10 PeV, near r=5 m). The difference of neutrons is the largest, the difference is 10\%-20\% at r$>$100m and the maximum difference is about 40\% (corresponding to $r<10m$). Overall, the difference between the muon and neutron models is significantly reduced at r$>$100m. In the experiment, secondary particles larger than 100 m were selected for reconstruction. It can reduce the dependence of the strong interaction model effectively.
\par
For the incidence of 45° zenith angle, compared with the vertical incidence, the number of electromagnetic particles will be significantly reduced and the fluctuation will be larger and the energy reconstruction accuracy and particle identification ability obtained by them will be worse. The number of muses will be slightly reduced and the fluctuation of the number of muses will not change much. The detection performance will be little affected. The decrease in the number of neutrons and the increase in the number of particles is somewhere between an electromagnetic particle and a muon. The change of Cherenkov photon relative to the vertical incidence depends on the vertical distance between the photon and the shower axis and the fluctuation of the photon number is also larger relative to the vertical incidence. And the amplitude of the fluctuation is similar to that of electromagnetic particles. In summary, when the atmospheric depth exceeds the atmospheric depth where the shower develops to the maximum, the fluctuation of various secondary components increases and the detection performance deteriorates. The fluctuation of the particle changes the least and the fluctuation of the electromagnetic particle changes the most.
\par
In summary, without considering the detector effect, this paper studies the energy reconstruction accuracy of various secondary components during energy reconstruction and the identification ability of the original particle composition, which provides references for the selection of detector types, energy reconstruction variables and methods and component identification variables in ground experiments.

\section*{Acknowledgements}
This work is supported by the Science and Technology Department of Sichuan Province, China (Grant No. 2021YFSY0031, 2020YFSY0016), the National Key R\&D Program of China (Grant No. 2018YFA0404201), and the National Natural Science Foundation of China (Grant Nos. 12205244, 12147208).

$\,$

$\,$

\end{CJK*}
\end{document}